\documentclass[aps,pra,reprint,superscriptaddress,twocolumn]{revtex4}
\usepackage[dvipdfmx]{graphicx}
\usepackage{graphicx}
\usepackage{color}
\usepackage[T1]{fontenc}
\usepackage{textcomp}
\usepackage{bm}
\usepackage{amsmath,amssymb,amsthm}
\usepackage{comment}
\usepackage{ulem}
\usepackage{grffile}
\usepackage{algorithm}
\usepackage{algpseudocode}

\newcommand{\cket}[1]{\left|#1\right\rangle}

\begin{document}


\title{Performance evaluation of the discrete truncated Wigner approximation for quench dynamics of quantum spin systems with long-range interactions}

\author{Masaya Kunimi}
\email{kunimi@ims.ac.jp}
\affiliation{Department of Photo-Molecular Science, Institute for Molecular Science, National Institutes of Natural Sciences, Myodaiji, Okazaki 444-8585, Japan}
\author{Kazuma Nagao}
\affiliation{Zentrum f\"{u}r Optische Quantentechnologien and Institut f\"{u}r Laserphysik,
Universit\"{a}t Hamburg, 22761 Hamburg, Germany}
\affiliation{The Hamburg Centre for Ultrafast Imaging, Luruper Chaussee 149, 
22761 Hamburg, Germany}
\author{Shimpei Goto}
\author{Ippei Danshita}
\affiliation{Department of Physics, Kindai University, Higashi-Osaka, Osaka 577-8502, Japan}


\date{\today}

\begin{abstract}
The discrete truncated Wigner approximation (DTWA) is a powerful tool for analyzing dynamics of quantum spin systems. Since the DTWA includes the leading-order quantum corrections to a mean-field approximation, it is naturally expected that the DTWA becomes more accurate when the range of interactions of the system increases. However, quantitative corroboration of this expectation is still lacking mainly because it is generally difficult in a large system to evaluate a timescale on which the DTWA is quantitatively valid. In order to investigate how the validity timescale depends on the interaction range, we analyze dynamics of quantum spin models with a step function type interaction subjected to a sudden quench of a magnetic field by means of both DTWA and its extension including the second-order correction, which is derived from the Bogoliubov-Born-Green-Kirkwood-Yvon equation. We also develop a formulation for calculating the second-order R\'enyi entropy within the framework of the DTWA. By comparing the time evolution of the R\'enyi entropy computed by the DTWA with that by the extension including the correction, we find that both in the one- and two-dimensional systems the validity timescale increases algebraically with the range of the step function type interaction.
\end{abstract}

\maketitle

\section{Introduction}\label{sec:Introduction}

The state-of-the-art technologies established in ultracold gases have opened the door for controlling and exploring coherent quantum dynamics of isolated many-body systems both near and far from equilibrium \cite{Bloch2012}. In the context of condensed-matter and solid-state physics, charge-neutral atoms loaded onto an optical lattice have been extensively studied as an analog quantum simulator for the tight-binding Hubbard-type models with short-range interactions. Owing to its controllability and cleanness, one can gain access to fundamental questions about dynamical properties of Hubbard-type systems.  The recent topics explored in experiments include thermalization dynamics of an isolated quantum system~\cite{Trotzky2012,Gring2012}, propagation of non-local correlations~\cite{Cheneau2012,Takasu2020a}, the Kibble--Zurek mechanism across quantum phase transitions \cite{Braun2015}, and the many-body localization (MBL) in a disordered optical lattice \cite{Choi2016,Abanin2019}. In recent years, technological developments in creating, controlling, and probing cold atoms or molecules with strong dipole-dipole interactions in an optical lattice \cite{dePaz2013,Yan2013,Hazzard2014,dePaz2016,Lepoutre2019,Fersterer2019,Patscheider2020}, Rydberg gases \cite{Schauss2015,Zeiher2016,Labuhn2016,Haenel2017,Zeiher2017,Bernien2017,Sous2018,Leseleuc2018,Guardado-Sanchez2018,Sous2019,Leseleuc2019,Keesling2019,Mizoguchi2019a}, and trapped ions \cite{Britton2012,Islam2013,Jurcevic2014,Richerme2014,Bohnet2016,Garttner2017} have enabled quantum simulation of various quantum spin systems with long-range interactions. In particular, Rydberg gases can be manipulated by means of the optical tweezer techniques, so that these offer an intriguing playground to explore novel quantum magnetism and non-equilibrium dynamics of localized spins caused by variable-range interactions.

While these experimental advances have stimulated theoretical studies of quantum many-body dynamics of systems with various interaction ranges, they are still limited due to the lack of available computational tools. As a quasi-exact numerical method, the time-dependent density-matrix renormalization group (tDMRG) has been typically utilized for simulating large-scale many-body systems corresponding to actual experiments \cite{vidal_efficient_2003,white_real-time_2004}. However, its efficient applications are limited to one-dimensional (1D) systems. Among various candidates for approximate frameworks to tackle many-body systems, the phase-space methods, especially the truncated-Wigner approximation (TWA) \cite{Blakie2008,Polkovnikov2010} on the basis of the Wigner--Weyl correspondence, provide a realistic and widely applicable approach to quantum many-body dynamics even for higher-dimensional systems with long-range interactions \cite{Tuchman2006,Davidson_thesis,Nagao2019,Ruostekoski2005,Fujimoto2019,Davidson2017}. Employing the TWA, quantum dynamics are reduced to a semiclassical problem of simulating randomly distributed classical trajectories in a phase space, each of which obeys a saddle-point or mean-field equation of motion for a given quantum system. The TWA gives quantitative descriptions of quantum dynamics even at long times if the system is in a certain classical limit or nearly non-interacting limit. More precisely, the TWA is asymptotically exact at short times \cite{Polkovnikov2003}. It implies that, within the TWA, there exists a threshold timescale separating semiclassically simulatable and non-simulatable regimes of quantum dynamics depending on the choice of the phase space. As demonstrated in some works \cite{Davidson2015,Nagao2020a,Wurtz2018}, by increasing the number of phase-space variables, one can improve the validity of TWA descriptions for strongly correlated lattice systems composed of bosons or spins. Such an increased phase-space approach is referred to as the SU($N$) or cluster TWA. Furthermore, it is also possible to construct a fermionic TWA (fTWA) approach for interacting fermions, in which $so(2N)$ string variables are introduced for a fermionic mode number $N$ \cite{Davidson_thesis,Davidson2017}. The fTWA has been used to study semiclassical aspects of chaos in the Sachdev--Ye--Kitaev (SYK) model \cite{Schmitt2019,Scaffidi2019}, which consists of all-to-all (infinitely long-range) two-body hoppings.

For describing dynamics of quantum spin systems, the discrete TWA (DTWA) has been widely applied in various contexts \cite{Schachenmayer2015,Schachenmayer2015_2,Babadi2015,Pucci2016,Orioli2017,Acevedo2017,Covey2018,Mori2019,Qu2019,Morong2020a,Orioli2018,Czischek2018,Sundar2019,Signoles2019a,Lepoutre2019,Fersterer2019,Patscheider2020}. In the conventional use of the TWA for spin systems \cite{Davidson2015,Polkovnikov2010,Wurtz2018}, to be efficient, its Monte Carlo sampling part employs a Gaussian approximation for the continuous Wigner distribution function. On the other hand, the DTWA utilizes a discrete Wigner function for sampling phase-space variables instead of the continuous Wigner distribution. Since the discrete Wigner representation is defined for the basis of local-spin eigenstates rather than coherent states, it can express typical initial states such as the all down-spin state $|\downarrow\downarrow\downarrow\downarrow\cdots\rangle$ and the staggered magnetization state $|\uparrow\downarrow\uparrow\downarrow\cdots\rangle$ as a positive-valued distribution. Thanks to this advantage, the DTWA accurately describes all the initial moments of these states and can capture some revival properties of quantum dynamics beyond the Gaussian approximation. These properties have been investigated in Ref.~\cite{Sundar2019}. More interestingly, the DTWA can also reproduce the experimental results for Rydberg atoms \cite{Orioli2018,Signoles2019a} and dipolar atoms \cite{Lepoutre2019,Fersterer2019,Patscheider2020}, which are effectively described by spin-$\frac{1}{2}$ and large-$S$ models, respectively.

Although the DTWA is a powerful tool to analyze quantum spin systems, there are some problems. One is a timescale on which the DTWA is valid. Generally, the TWA framework gives quantitatively valid results in a short time regime \cite{Blakie2008,Polkovnikov2010}. This validity timescale depends on the details of the systems, such as interactions, dimensions, and initial conditions. In the zero-temperature ground states or thermal equilibrium cases, it is well established that the mean-field approximation gives the exact results in large-$S$ limit or infinite dimensions (or all-to-all coupling). As for the quantum dynamics, it has been shown that the validity timescale becomes longer when the size of the spin $S$ increases~\cite{Zhu2019}. By contrast, the dependence of the validity timescale on the spatial dependence of the interaction is not well understood. In Refs.~\cite{Schachenmayer2015,Schachenmayer2015_2,Sundar2019}, the performance of the DTWA for power-law decay interaction has been discussed. Another possibility of interaction form is a step function type interaction, which approximately describes the spin-spin interaction of the Rydberg-dressed atoms \cite{Zeiher2016,Zeiher2017}. Investigating such a range-dependence of the validity timescale will be useful when the TWA is applied to analyzing such systems, in which the interaction range can be systematically controlled.

In this paper, we address the question how the validity timescale of the DTWA depends on the interaction range of the step function type interaction. Our approach is to use higher-order corrections of the DTWA, which are derived by using the Bogoliubov-Born-Green-Kirkwood-Yvon (BBGKY) hierarchy equation \cite{Pucci2016}. When the difference between the DTWA and its higher-order corrections is small, we can expect that the DTWA is a good approximation. 

In order to compare the DTWA with its higher-order corrections, we focus on the second-order R\'enyi entanglement entropy. In particular, we consider two-site R\'enyi entanglement entropy, which contains the information of one- and two-spin expectation values. In this paper, we develop a method to calculate the R\'enyi entropy within the framework of the DTWA. Our method can be applicable to not only the benchmark of the DTWA, but also the calculations of the R\'enyi entropy in higher dimensions, which are difficult to access by other methods. 

From the comparison between the DTWA and its extension including higher-order corrections, we show that the validity timescale of the DTWA becomes longer as increasing the range of the step function type interaction. We confirm this property for three different kinds of quantum-spin models, namely, Ising, XY, and Heisenberg model under a uniform magnetic field in 1D and 2D. This result means that the DTWA becomes better as the classical limit is approached.

This paper is organized as follows: In Sec.~\ref{sec:model}, we explain our model and the DTWA. In Sec.~\ref{subsec:criterion}, we explain how to define the threshold time on which the DTWA is valid and compare the DTWA and tDMRG results. In Sec.~\ref{subsec:renyi_and_threshold}, we show the results of R\'{e}nyi entanglement entropy and threshold time for three different quantum-spin models. In Sec.~\ref{subsec:compare1D_and_2D}, we compare the results in 1D and 2D systems. In Sec.~\ref{sec:summary}, we summarize our results. In Appendix \ref{sec:details_of_DTWA}, we discuss the details of the derivation of the DTWA.  In Appendix \ref{sec:details_of_DTWA_sampling}, we explain the sampling scheme of the initial conditions.  In Appendix \ref{sec:derivation_of_Renyi_entropy}, we derive the expression of the R\'enyi entanglement entropy in the framework of the DTWA. In Appendix \ref{sec:comparison_with_DTWA_and_tDMRG}, we explain the details of the tDMRG calculations. In Appendix \ref{sec:new_algorithm}, we propose an algorithm to calculate the dynamics of the systems with the long-range interaction by using the tDMRG.

\section{Model and methods}\label{sec:model}
\subsection{Model}\label{subsec:model}

In this paper, we consider a family of quantum spin-$\frac{1}{2}$ systems in 1D and 2D, which is generally modeled by the Hamiltonian 
\begin{align}
\hat{H}=\frac{1}{2}\sum_{i,j, i \not=j}\sum_\mu J_{i j}^{\mu}\hat{S}_i^{\mu}\hat{S}_j^{\mu}+\bm{h}\cdot\sum_i\hat{\bm{S}}_i,\label{eq:system_Hamiltonian}
\end{align}
where $\hat{S}_i^{\mu}\; (\mu=x,y,z)$ is a spin-$\frac{1}{2}$ operator at site $i$, $J_{i j}^{\mu}$ is a  spin-exchange coupling between two distant sites, and $\bm{h}\equiv (h^x, h^y, h^z)$ is a uniform magnetic field, respectively. Throughout this paper, we impose open boundary conditions and write $M$ as an even integer expressing the total number of lattice sites.  

As a concrete form of the coupling $J_{i j}^{\mu}$, we especially focus on three specific cases as follows:
\begin{align}
&J_{ij}^x=J_{ij}^y=0,\quad J_{ij}^z=-J_{i j}^{(D)},\quad (\text{Ising}),\label{eq:definition_of_Ising_interaction_1D}\\
&J_{ij}^x=J_{ij}^y=-J_{i j}^{(D)},\quad J_{ij}^z=0,\quad(\text{XY}),\label{eq:definition_of_XY_interaction_1D}\\
&J_{ij}^x=J_{ij}^y=J_{ij}^z=J_{ij}^{(D)},\quad(\text{Heisenberg}).\label{eq:definition_of_Heisenberg_interaction_1D}
\end{align}
From the top to bottom, let us refer to these as a ferromagnetic Ising, ferromagnetic XY, and antiferromagnetic Heisenberg models, respectively. Furthermore, we assume that a magnetic field is applied along the $x$ axis, i.e., $\bm{h} = (h^x, 0, 0)$. The details of $J^{(D)}_{ij}$ depend on the spatial dimension of the lattice $D$. For the 1D cases ($D=1$), it has the properties
\begin{align}
J_{ij}^{(1)}\equiv
\begin{cases}
\vspace{0.2em}\dfrac{J}{r},\quad \text{if } |i-j|\le r,\\
0,\quad \text{otherwise}
\end{cases}
\label{eq:definition_1D_spin-spin_interaction}
\end{align}
where $J>0$ is an interaction strength and $r=1,2,\cdots$ is the interaction range. By contrast, the 2D cases ($D=2$), in which the lattice geometry is supposed to be square, are characterized by 
\begin{align}
J_{ij}^{(2)}\equiv 
\begin{cases}
\vspace{0.2em}\dfrac{J}{C_r},\quad \text{if }|\bm{R}_i-\bm{R}_j|\le R_r,\\
0,\quad\text{otherwise}
\end{cases}
\label{eq:definition_of_2D_spin-spin_interaction}
\end{align}
where $\bm{R}_i\equiv (R_{x i}, R_{y i})\equiv a(i_x, i_y)$ is the position of the $i$th lattice site, $a$ is the lattice constant, $i_x=1,2,\cdots, M_x$ and $i_y=1,2,\cdots, M_y$ are indices of the $i$th lattice sites, and $R_r$ is the distance between the $r$th neighboring sites. The total number of lattice points is given by $M=M_xM_y$. The constant $C_r$ is determined such that the following equation is satisfied:
\begin{align}
\max_i\sum_{j\not=i}J_{i j}^{(2)}=2J.\label{eq:definition_of_constant_Cr}
\end{align}
The explicit values of $C_r$ are given by $C_1=2, C_2=4, C_3=6, C_4=10, C_5=12,\cdots$. For later use, we define $N_r$ as
\begin{align}
N_{r}=
\begin{cases}
2r,\quad (\text{1D}),\\
2C_r,\quad (\text{2D}),
\end{cases}
\label{eq:definition_of_Nr}
\end{align}
where $N_r$ approximately denotes the number of connections per spin quantifying how many spins are connected to each spin. Approaching the boundaries from the center of the system, the actual number of connections decreases from $N_r$ due to the finite system size and the open boundary.

It is worth noting that these types of long-range interaction, Eqs.~(\ref{eq:definition_1D_spin-spin_interaction}) and (\ref{eq:definition_of_2D_spin-spin_interaction}), are realizable in the experimental setups by means of Rydberg-dressed atoms \cite{Zeiher2016,Zeiher2017}. A system of such atoms is typically characterized by a soft-core type potential, so that interactions among atoms are almost constant in the short-distance regime and rapidly decay in the long-distance regime \cite{Henkel2010,Pupillo2010}. The couplings in Eqs.~(\ref{eq:definition_1D_spin-spin_interaction}) and (\ref{eq:definition_of_2D_spin-spin_interaction}) may describe such a situation approximately. 

In the subsequent sections we will investigate sudden-quench dynamics of the interacting spin systems \cite{Hauke2013,Schachenmayer2013,Lepori2016,Buyskikh2016} in order to characterize the limitation of the DTWA method. To be concrete, we especially consider the following direct-product wave functions as low-entangled initial states:
\begin{align}
|\psi(0)\rangle&=\prod_{i=1}^M|\leftarrow_i\rangle,\quad (\text{Ising}),\label{eq:initial_state_for_Ising_model}\\
|\psi(0)\rangle&=\prod_{i=1}^M|\downarrow_i\rangle,\quad (\text{XY}),\label{eq:initial_state_for_XY_model}\\
|\psi(0)\rangle&=\prod_{i=1}^{M/2}|\uparrow_{2i-1}\downarrow_{2i}\rangle,\quad (\text{Heisenberg}).\label{eq:initial_state_for_Heisenberg_model}
\end{align}
Here $\cket{\uparrow_i}$ and $\cket{\downarrow_i}$ denote the eigenstates of $\hat{S}_i^z$ while $\cket{\rightarrow_i}\equiv (\cket{\uparrow_i}+\cket{\downarrow_i})/\sqrt{2}$ and $\cket{\leftarrow_i}\equiv (\cket{\uparrow_i}-\cket{\downarrow_i})/\sqrt{2}$ represent the ones of $\hat{S}_i^x$. The corresponding discrete Wigner functions for these initial states are shown in Appendix~\ref{sec:details_of_DTWA_sampling}.

\subsection{Discrete phase-space approach to the R\'enyi entropy}\label{subsec:DTWA}

In this work, to characterize the performance of the DTWA, we focus on the second-order R\'enyi entanglement entropy defined by
\begin{align}
S^{(2)}_A(t)\equiv -\log\left({\rm Tr}\left\{[\hat{\rho}_A(t)]^2\right\}\right),\label{eq:definition_of_second_order_Renyi_entropy}
\end{align}
where $\hat{\rho}_A(t)\equiv {\rm Tr}_B\hat{\rho}(t)$ is the reduced density matrix associated with a subregion $A$ and $\hat{\rho}(t)$ is the density matrix of the whole system. The whole system is separated in real space into $A$ and $B$. In what follows, let us derive a discrete phase-space representation for $S^{(2)}_A(t)$.

An important ingredient to make the discrete phase-space representation is the phase-point operator ${\hat A}_{\bm{\alpha}}$ \cite{Schachenmayer2015}. 
For SU(2) spin systems, it generally takes the form
\begin{align}
\hat{A}_{\bm{\alpha}}\equiv \prod_{i=1}^M\left(\frac{1}{2}+\bm{r}_{\alpha_i}\cdot\hat{\bm{S}}_i\right), \label{eq:defintion_of_phase-point_operator}
\end{align}
where $\bm{\alpha}\equiv (\alpha_1,\alpha_2,\cdots,\alpha_M)$ with $\alpha_i \in \{(0,0), (0, 1), (1, 0), (1,1)\}$ denotes independent points in the discrete phase space and $\bm{r}_{\alpha_i}$ is a three-dimensional vector implying $\bm{r}_{(0,0)} = (+1,+1,+1)$, $\bm{r}_{(0,1)} = (-1,-1,+1)$, $\bm{r}_{(1,0)} = (+1, -1, -1)$, and $\bm{r}_{(1,1)} = (-1,+1,-1)$ \cite{Wootters1987}.
The density matrix at time $t$ can be written as \cite{Wootters1987}
\begin{align}
\hat{\rho}(t)=\sum_{\bm{\alpha}}W_{\bm{\alpha}}(0)\hat{A}_{\bm{\alpha}}(t),\label{eq:density_matrix_expressed_by_phase_point_operator}
\end{align}
where $W_{\bm{\alpha}}(0)$ is the discrete Wigner function at $t=0$ and we defined $\hat{A}_{\bm{\alpha}}(t)=e^{-i\hat{H}t/\hbar}\hat{A}_{\bm{\alpha}}e^{+i\hat{H}t/\hbar}$. The reduced density matrix $\hat{\rho}_A(t)$ is expressed by means of the discrete Wigner function at $t=0$ and $\hat{A}_{\bm{\alpha}}(t)$:
\begin{align}
\hat{\rho}_A(t)={\rm Tr}_B \left\{ \sum_{\bm{\alpha}}W_{\bm{\alpha}}(0)\hat{A}_{\bm{\alpha}}(t) \right\}.\label{eq:reduced_density_matrix_by_Discrete_Wigner_function}
\end{align}

The transformed phase-point operator $\hat{A}_{\bm{\alpha}}(t)$ contains a complete information about quantum many-body dynamics governed by ${\hat H}$. However, carrying out exact calculations for such an operator is generally impossible.  

In the descriptions of DTWA, the phase-point operator at $t$ is assumed to be factorized \cite{Schachenmayer2015}, i.e., 
\begin{align}
\hat{A}_{\bm{\alpha}}(t)\simeq \prod_{i=1}^M\left[ \frac{1}{2}+\bm{r}_i(t,\bm{\alpha})\cdot\hat{\bm{S}}_i \right]. \label{eq:ppo_dtwa}
\end{align}
The time-dependent coefficient $\bm{r}_i(t,\bm{\alpha}) \equiv 2\bm{S}_i(t)$ obeys a classical equation of motion obtained from a first-order BBGKY hierarchy truncation (see also Appendix~\ref{sec:details_of_DTWA})
\begin{align}
\hbar\frac{d}{d t}S_i^{\mu}(t)&=\epsilon_{\mu\beta\gamma}\left[h^{\beta}S_i^{\gamma}(t)+\sum_{k\not=i}J_{i k}^{\beta}S_k^{\beta}(t)S_i^{\gamma}(t)\right],\label{eq:classical_equation_of_motion_DTWA_1st}
\end{align}
where $\epsilon_{\mu\beta\gamma}$ is the Levi-Civita symbol and we used the Einstein notation for repeated Greek indices. 
Inserting Eq.~(\ref{eq:ppo_dtwa}) into Eq.~(\ref{eq:definition_of_second_order_Renyi_entropy}), we arrive at a DTWA expression of the R\'enyi entropy
\begin{align}
S_{A}^{(2)}(t) \approx -\log\left\langle\left\langle\prod_{i\in A}\left[\frac{1}{2}+2\bm{S}_i(t)\cdot\bm{S}'_i(t)\right]\right\rangle\right\rangle, \label{eq:1st_order_BBGKY_renyi_expression}
\end{align}
where $\bm{S}_i(0)=\bm{r}_{\alpha_i}/2$ and $\bm{S}'_i(0)=\bm{r}_{\alpha_i'}/2$. The doubled angular brackets mean a phase-space average weighted with two initial Wigner functions 
\begin{align}
\left\langle\left\langle f_{\bm{\alpha},\bm{\alpha}'} \right\rangle\right\rangle \equiv \sum_{\bm{\alpha},\bm{\alpha}'}W_{\bm{\alpha}}(0)W_{\bm{\alpha}'}(0)f_{\bm{\alpha},\bm{\alpha}'}.
\end{align}

For direct product states such as Eqs.~(\ref{eq:initial_state_for_Ising_model}), (\ref{eq:initial_state_for_XY_model}), and (\ref{eq:initial_state_for_Heisenberg_model}), $W_{\bm{\alpha}}(0)$ is factorized as $W_{\bm{\alpha}}(0)=\prod_{j=1}^{M}w_{\alpha_{j}}(0)$.  It means that each local spin variable can fluctuate independently and the entropy for a subsystem results in zero at $t=0$. The subsystem entropy remains zero during the time evolution if the Hamiltonian is entirely decoupled into local parts and each part is linear in SU(2) matrices. For nonlinear systems with a nonzero spin-exchange coupling, the R\'enyi entropy becomes nonzero as a consequence of the many-body time evolution. The semiclassical expression for the subsystem entropy states that the amount of entanglement across the boundary of two subregions is related to the degree of {\it complexity} in a solution of Eq.~(\ref{eq:classical_equation_of_motion_DTWA_1st}) that is, if it is possible to write down, provided as a complicated non-linear function of initial conditions.

A higher-order correction beyond the DTWA description based on Eq.~(\ref{eq:classical_equation_of_motion_DTWA_1st}) arises in a classical trajectory of an enlarged phase space for a second-order BBGKY method \cite{Pucci2016}. The underlying idea of this method is to regard a nonseparable part of ${\hat S}^{\mu}_{i}{\hat S}^{\nu}_{j}$, which is represented by $c^{\mu\nu}_{ij}$ in a replacement ${\hat S}^{\mu}_{i}{\hat S}^{\nu}_{j} \rightarrow S_i^{\mu}S_j^{\nu}+c_{ij}^{\mu\nu}$, as an additional mechanical variable and define an approximately closed equation of motion for $S^{\mu}_{i}$ and $c_{ij}^{\mu\nu}$. We note that the first-order BBGKY truncation leads to the time-evolving equation given by Eq.~(\ref{eq:classical_equation_of_motion_DTWA_1st}). In Sec.~\ref{sec:results}, we will also exploit the second-order BBGKY method to compute the R\'enyi entropy, especially for a subregion of two sites. The detail of the BBGKY formulation will be presented in Appendix \ref{sec:details_of_DTWA}.

In this paper, we numerically solve Eq.~(\ref{eq:classical_equation_of_motion_DTWA_1st}) for the first-order BBGKY and  Eqs.~(\ref{eq:BBGKY_first_order_equation_re}) and (\ref{eq:BBGKY_second_order_re}) for the second-order BBGKY by using a fourth-order Runge-Kutta method. A time step $\Delta t$ is taken to be $\Delta t=10^{-3}\hbar/J$. We have checked that our results are converged for this $\Delta t$ in our simulation timescale. We use $M=100$ in 1D and $M_x=M_y=14$ in 2D. The computational costs for first- and second-order BBGKY equation scale as $O(M)$ and $O(M^2)$, respectively.

\section{Results}\label{sec:results}

\subsection{Criterion for the validity of the DTWA and comparison with tDMRG results}\label{subsec:criterion}

In this subsection, we introduce a criterion for giving an estimation of the timescale within which the DTWA is quantitatively valid. Our approach is based on the assumption that when the difference between the first- and second-order BBGKY results is small, the DTWA gives a good approximation. Here, a question arises: Which physical quantities are appropriate for comparing the first- and second-order BBGKY results? We propose that the second-order R\'{e}nyi entanglement entropy is a suitable quantity for confirming the validity of the DTWA. One advantage to use the R\'enyi entropy is that it is an unbiased quantity compared with other physical quantities such as spin expectation values and spin-spin correlations. The latter quantities strongly depend on the dynamics and symmetry of the systems. For example, if the system has spin rotational symmetry along the $z$ axis, $\hat{S}_{\rm tot}^z\equiv \sum_i\hat{S}_i^z$ is conserved so that it is not appropriate for examining the validity of the DTWA.

In this paper, we calculate mean two-site R\'enyi entropies for 1D and 2D, which are defined by
\begin{align}
S^{(2)}(t)\equiv \frac{1}{M-1}\sum_{i<j,|i-j|=1}S_{ij}^{(2)}(t),\quad (\text{1D}),\label{eq:definition_mean_Renyi_1D}\\
S^{(2)}(t)\equiv \frac{1}{(M_x-1)M_y}\sum_{i<j,|R_{xi}-R_{xj}|=a}S_{ij}^{(2)}(t),\quad (\text{2D}).\label{eq:definition_mean_Renyi_2D}
\end{align}
where $S_{ij}^{(2)}(t)$ is the two-site R\'enyi entropy. Let us mention differences between our formulation of the R\'{e}nyi entropy and that of previous works \cite{Acevedo2017,Lepoutre2019,Zhu2019}. In these previous works, they calculated the single- or two-site R\'{e}nyi entropy from the expressions of the single- or two-site reduced density-matrix operator because the matrix elements of these reduced density matrices can be constructed by the expectation values of $\hat{S}_i^{\mu}$ and $\hat{S}_i^{\mu}\hat{S}_j^{\nu}$, which can be obtained by the DTWA. The advantage of our formulation is that it allows us to calculate the R\'{e}nyi entropy for multiple sites. It is easy to calculate the multiple-sites R\'{e}nyi entropy in the first-order BBGKY. We have checked the statistical convergence of the results even in the case of bipartite R\'enyi entropy. For more details, see Appendix \ref{sec:derivation_of_Renyi_entropy}.  Here, we focus on the case where the two sites are nearest neighbors in 1D. In 2D, we consider the neighboring sites of the $x$ direction only. 

We calculate the two-site R\'enyi entropy by using the first- and second-order BBGKY equations. In order to quantify the difference between the first- and second-order results, we define
\begin{align}
\Delta(t)\equiv \left|e^{-S_{\rm 1st}^{(2)}(t)}-e^{-S_{\rm 2nd}^{(2)}(t)}\right|/e^{-S_{\rm 1st}^{(2)}(t)},
\end{align}
where $S_{\rm 1st}^{(2)}(t)$ and $S_{\rm 2nd}^{(2)}(t)$ are the mean two-site R\'enyi entropy obtained by the first- and second-order BBGKY equation, respectively. This quantity represents a relative error of the first- and second-order results. The reason why we do not use the relative error of $S_{\rm 1st}^{(2)}(t)$ and $S_{\rm 2nd}^{(2)}(t)$ is to avoid the divergence of the relative error because $S^{(2)}(0)=0$ in our initial conditions. Comparing these quantities, we can define a threshold time $T_{\rm th}$, at which $\Delta(t)$ exceeds a small positive number $\epsilon$.
In this paper, we use $\epsilon=\frac{1}{10}$.

\begin{figure}[t]
\centering
\includegraphics[width=8.5cm,clip]{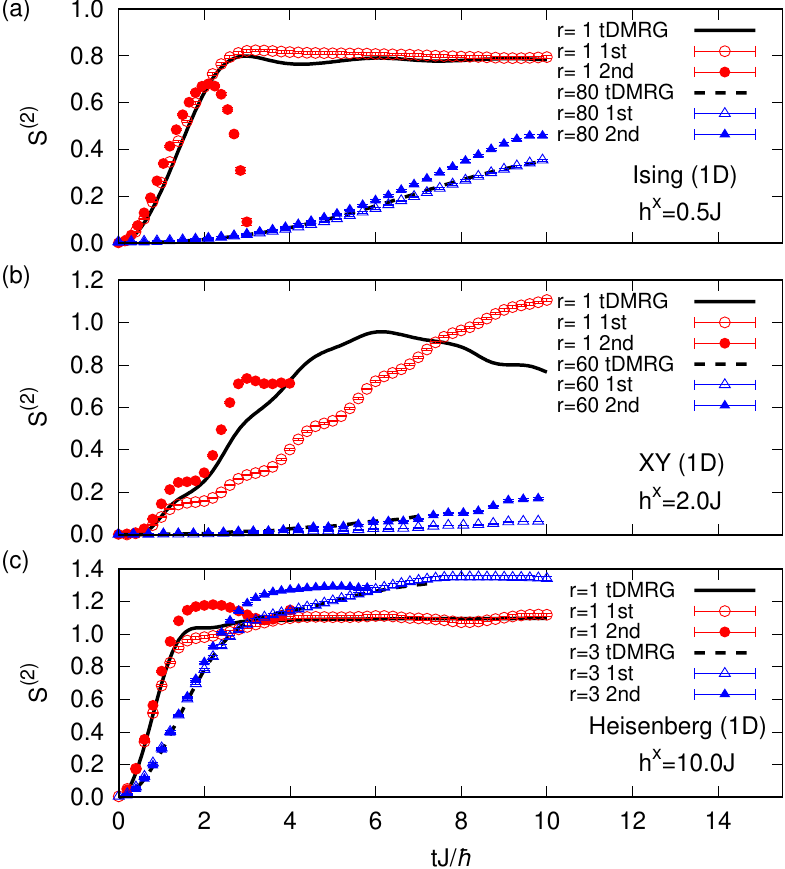}
\caption{Comparison with the DTWA and tDMRG results. Mean two-site R\'{e}nyi entropy in 1D. (a) Ising model for $h^x=0.5J$. (b) XY model for $h^x=2.0J$. (c) Heisenberg model for $h^x=10.0J$. The black solid and dotted lines represent the tDMRG results. The open and closed symbols represent the 1st and 2nd order BBGKY results, respectively. $r$ is the interaction range.}
\label{fig:compare_renyi_entropy_model_1D}
\end{figure}

We note that the second-order BBGKY equation is numerically unstable as pointed out in Refs.~\cite{Orioli2017,Czischek2018}. In fact, we find the divergent behavior of the second-order BBGKY equation. For example, this behavior can be seen in $r=1$ results in Figs.~\ref{fig:compare_renyi_entropy_model_1D} and \ref{fig:renyi_entropy_1D_and_2D}. This is an intrinsic property of the second-order BBGKY equation. We have checked that this divergence behavior is not an artificial one because it does not depend on the choice of the time step $\Delta t$.

The small difference between the first- and second-order results is a necessary condition for the DTWA to be good approximation. This criterion is based on the assumption that the 1st and 2nd order results can approximate the exact results. Even if the difference is small, our criterion is meaningless if the DTWA cannot reproduce the exact results. To corroborate that our criterion indeed works, we perform the comparison between the DTWA results and the tDMRG method in the 1D cases. See details of the tDMRG calculations in Appendix \ref{sec:comparison_with_DTWA_and_tDMRG}. Figure \ref{fig:compare_renyi_entropy_model_1D} shows the results of the mean two-site R\'enyi entropy for (a) Ising model, (b) XY model, and (c) Heisenberg model, respectively. For long-range Hamiltonians, we implement tDMRG with utilizing swap gates as detailed in Appendix~\ref{sec:new_algorithm}. We can see that the DTWA results are good agreement with the tDMRG results in a short time scale ($t\sim \hbar/J$) for all cases. In the long-range interacting cases, the DTWA results quantitatively reproduce the tDMRG results even in the long-time scale ($t\sim 10\hbar/J$). However, in the long-time scale, the second-order BBGKY results for some parameters are divergent (We can see this from the fact that there is no data point for second-order BBGKY results in the long-time regime.). This means that our criterion does not work in the long-time scale. Therefore, we expect that our criteria for the performance of the DTWA work in the short- and intermediate-time scale. In the following, we focus on these timescales.

\subsection{R\'{e}nyi entropy and threshold time}\label{subsec:renyi_and_threshold}

In this subsection, we show the results of R\'enyi entropy and threshold time for Ising, XY, and Heisenberg models.  We consider three different spin models in order to indicate that the statement that the validity timescale of the DTWA increases with increasing the interaction range holds regardless of the integrability and symmetry. In 1D, the Ising model with transverse magnetic field is integrable while the XY and Heisenberg models under a uniform magnetic field are nonintegrable. The Ising and XY models have only discrete symmetry while the Heisenberg model has continuous spin-rotation symmetry around the magnetic field.

\subsubsection{Ising model}\label{subsubsec:ising}

\begin{figure*}[t]
\centering
\includegraphics[width=17.0cm,clip]{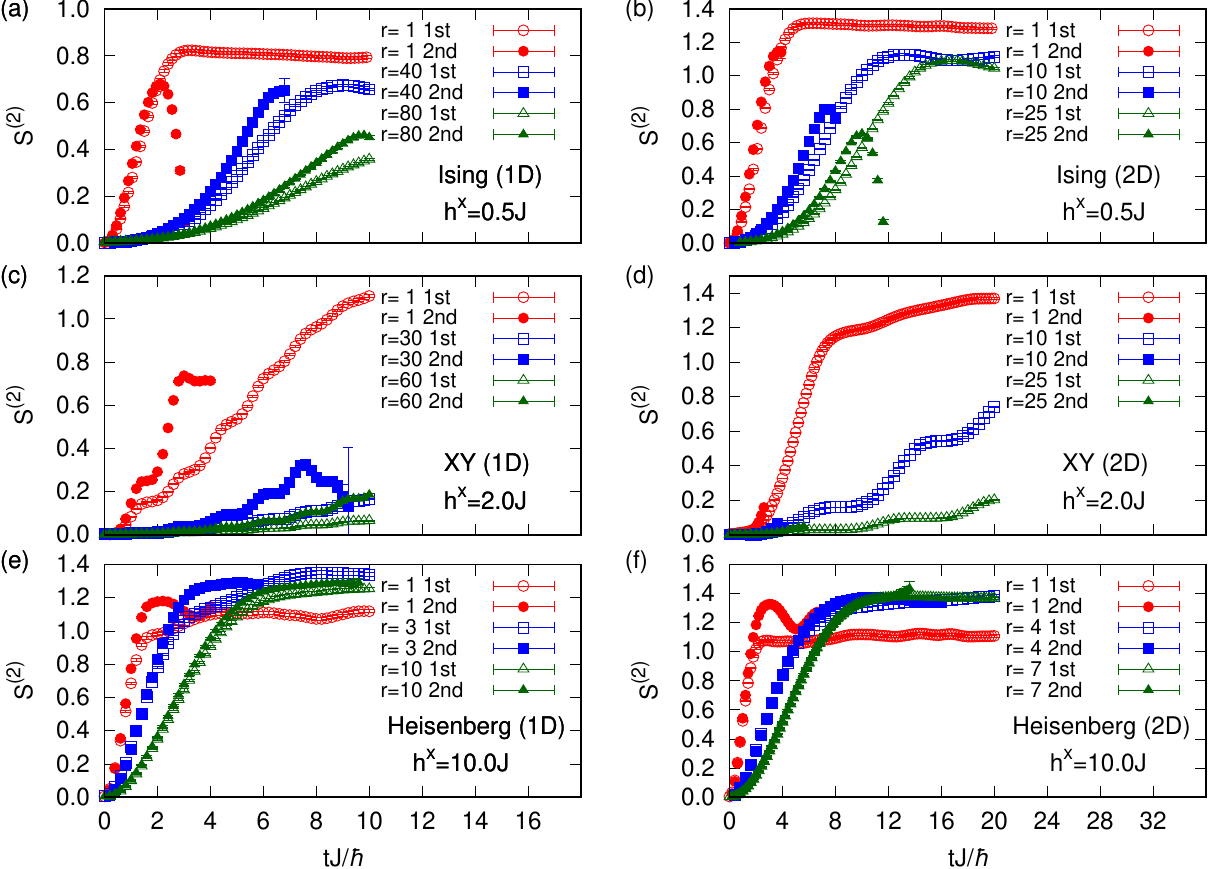}
\caption{Time evolution after the sudden quench of the mean two-site R\'{e}nyi entropy in 1D and 2D. (a), (b) Ising model for $h^x=0.5J$. (c), (d) XY model for $h^x=2.0J$. (e), (f) Heisenberg model for $h^x=10.0J$. The open and closed symbols represent the first- and second-order BBGKY results, respectively. $r$ denotes the interaction range.}
\label{fig:renyi_entropy_1D_and_2D}
\end{figure*}%

\begin{figure}[t]
\centering
\includegraphics[width=8.0cm,clip]{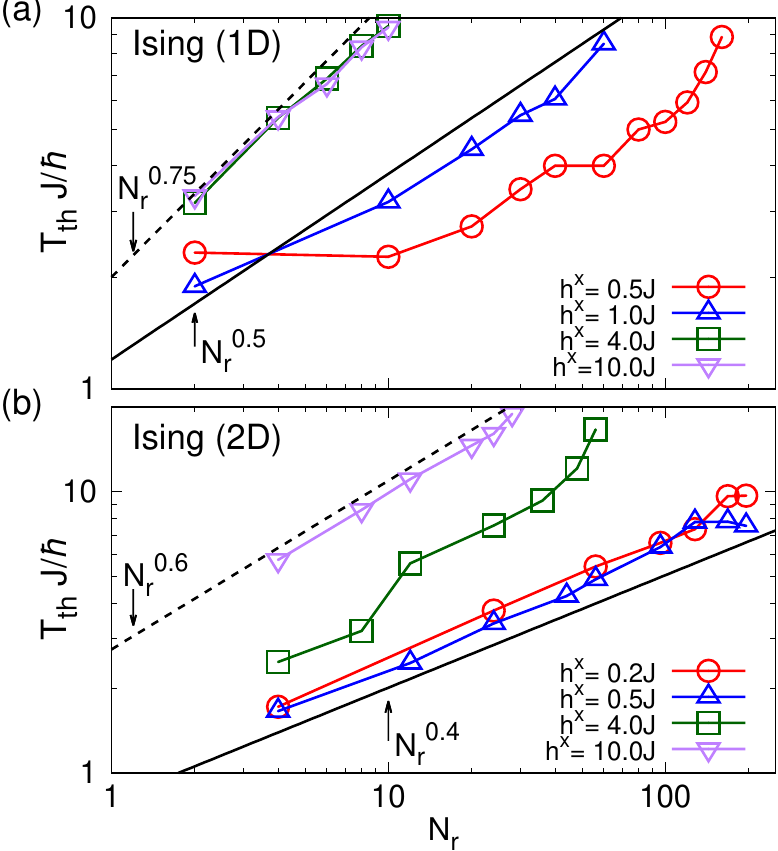}
\caption{Threshold time as a function of number of interacting sites for the Ising model. (a) One dimension. The black solid and dashed lines represent $N_r^{0.5}$ and $N_r^{0.75}$ for guide to eye. (b) Two dimensions. The black solid and dashed lines represent $N_r^{0.4}$ and $N_r^{0.6}$ for guide to eye.}
\label{fig:T_th_for_Ising_model}
\end{figure}%

Here, we show the results of the mean two-site R\'{e}nyi entropy for the Ising model under the transverse magnetic field in Fig.~\ref{fig:renyi_entropy_1D_and_2D} (a) for 1D and Fig.~\ref{fig:renyi_entropy_1D_and_2D} (b) for 2D. In these cases, the initial condition is the fully $-x$-polarized state [see Eq.~(\ref{eq:initial_state_for_Ising_model})]. This state is the exact ground state when $h^x\to \infty$. We can see the growth of the entanglement in an early stage of the dynamics. This behavior is a typical one in the quench dynamics of many-body systems \cite{Eisert2010}. In the long-range interacting systems, the growth of the entanglement is slow compared to the short-range interacting systems. This tendency is consistent with the previous work \cite{Schachenmayer2013}. In the $r\to\infty$ limit, our model becomes the Lipkin-Meshkov-Glick model \cite{Vidal2004,Latorre2005}. In this model, the dynamics is constructed by a small number of quantum states. The bipartite entanglement entropy is bounded by the logarithm of the system size. Although this property is related to the bipartite entanglement, we can naively expect that the two-site R\'enyi entropy has same tendency. 

From the results of the R\'enyi entropy, we can obtain the threshold time $T_{\rm th}$ as a function of $N_r$. The results are shown in Fig.~\ref{fig:T_th_for_Ising_model} (a) for 1D and Fig.~\ref{fig:T_th_for_Ising_model} (b) for 2D.  We can see that the threshold time increases as a power law of $N_r$ (see solid and dotted lines in Fig.~\ref{fig:T_th_for_Ising_model}. We can also see that $T_{\rm th}$ in the large-$h^x$ region is large compared to that in the small-$h^x$ region. This behavior can be understood from the fact that the DTWA yields exact results when interaction terms are absent. In the large-$h^x$ region, the dynamics is mainly driven by the magnetic field. In this reason, $T_{\rm th}$ in the large-$h^x$ region is longer than that of the small-$h^x$ region.

\subsubsection{XY model}\label{subsubsec:XY}

Next, we show the results of the XY model under the magnetic field along the $-x$ direction in Fig.~\ref{fig:renyi_entropy_1D_and_2D} (c) for 1D and \ref{fig:renyi_entropy_1D_and_2D} (d) for 2D. In this case, the initial condition is the fully $-z$-polarized state [see Eq.~(\ref{eq:initial_state_for_XY_model})]. This state is an exact eigenstate when $h^x=0$.

\begin{figure}[t]
\centering
\includegraphics[width=8.0cm,clip]{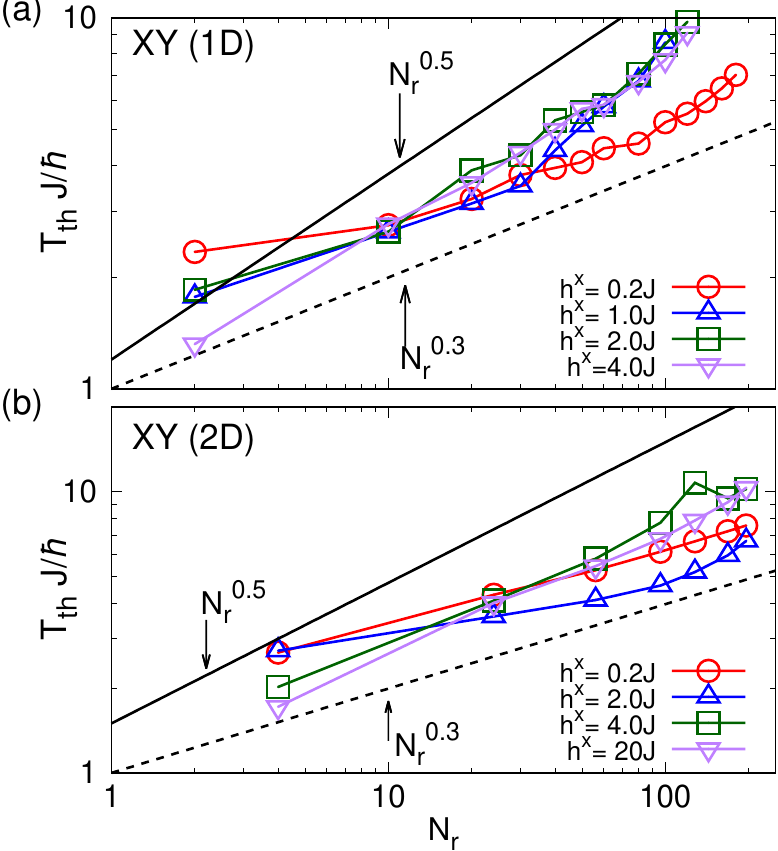}
\caption{Threshold time as a function of the number of interacting sites for the XY model. (a) One dimension. The black solid and dashed lines represent $N_r^{0.5}$ and $N_r^{0.3}$ for guide to eye. (b) Two dimensions. The black solid and dashed lines represent $N_r^{0.5}$ and $N_r^{0.3}$ for guide to eye.}
\label{fig:T_th_for_XY_model}
\end{figure}%

From these results, we obtain the threshold time as a function of $N_r$. The results are shown in Fig.~\ref{fig:T_th_for_XY_model} (a) for 1D and Fig.~\ref{fig:T_th_for_XY_model} (b) for 2D. The results are similar to those of the Ising model. We can see that the threshold time increases as power law of $N_r$ (see solid and dotted lines in Fig.~\ref{fig:T_th_for_XY_model}). The dependence of the magnetic field is also similar to that of the Ising model. The large-$h^x$ case is better than the small-$h^x$ case.

\subsubsection{Heisenberg model}\label{subsubsec:Heisenberg}

We show the results of the Heisenberg model under the magnetic field along the $-x$ direction in Fig.~\ref{fig:renyi_entropy_1D_and_2D} (e) for 1D and Fig.~\ref{fig:renyi_entropy_1D_and_2D} (f) for 2D. In this case, the initial condition is the Ne\'el state [see Eq.~(\ref{eq:initial_state_for_Heisenberg_model})]. In contrast to the previous cases, we consider the antiferromagnetic interaction and nonuniform initial condition. The reason is as follows. If we consider the ferromagnetic Heisenberg model with a fully-polarized initial condition under the uniform magnetic field, the resultant dynamics is the Larmor precession motion, which is not affected by the interaction. During this dynamics, the entanglement is exactly zero. Therefore, we need to consider another situation. 

\begin{figure}[t]
\centering
\includegraphics[width=8.0cm,clip]{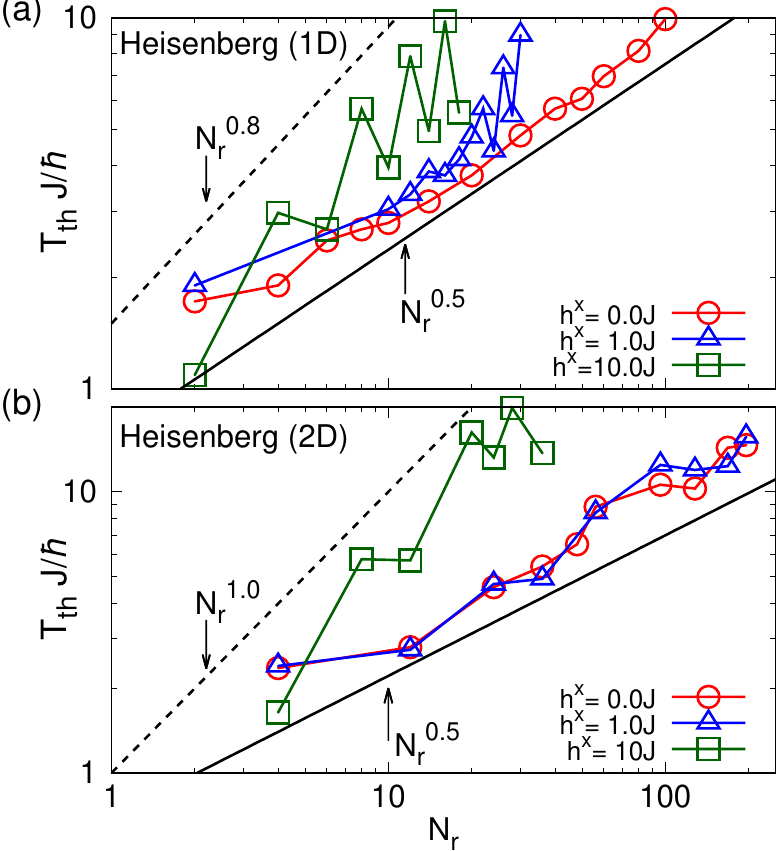}
\caption{Threshold time as a function of the number of interacting sites for the Heisenberg model. (a) One dimension. The black solid and dashed lines represent $N_r^{0.5}$ and $N_r^{0.8}$ for guide to eye. (b) Two dimensions. The black solid and dashed lines represent $N_r^{0.5}$ and $N_r^{1.0}$ for guide to eye.}
\label{fig:T_th_for_heisenberg_model}
\end{figure}%

The threshold time as a function of $N_r$ is in Fig.~\ref{fig:T_th_for_heisenberg_model}. We can see that the threshold time increases as power law of $N_r$ (see solid and dotted lines in Fig.~\ref{fig:T_th_for_heisenberg_model}). The dependence of the magnetic field is also similar to that of the Ising and XY models. In contrast to the previous two models, we can see an oscillation behavior of $T_{\rm th}$. For example, the oscillation of $h^x=10J$ shown in Fig.~\ref{fig:T_th_for_heisenberg_model} (a) can be seen clearly. This is due to the relation between the interaction and the initial condition. To explain this behavior, we consider $r=1$ and $2$ cases in an early-time regime. In the $r=1$ case, each pair of spins coupled via the Heisenberg interaction are aligned antiparallelly in the initial state such that the initial state has relatively low energy. However, in the $r=2$ case, the next-nearest-neighbor interactions couple parallelly aligned pairs of spins in the initial state such that the initial state has much higher energy than the $r=1$ case. In other words, the injected energy by the quench alternates when $r$ increases one by one. Therefore, the oscillating behavior of $T_{\rm th}$ as a function of the interaction range appears.

\subsection{Comparison with one- and two-dimensional results}\label{subsec:compare1D_and_2D}

\begin{figure}[t]
\centering
\includegraphics[width=8.5cm,clip]{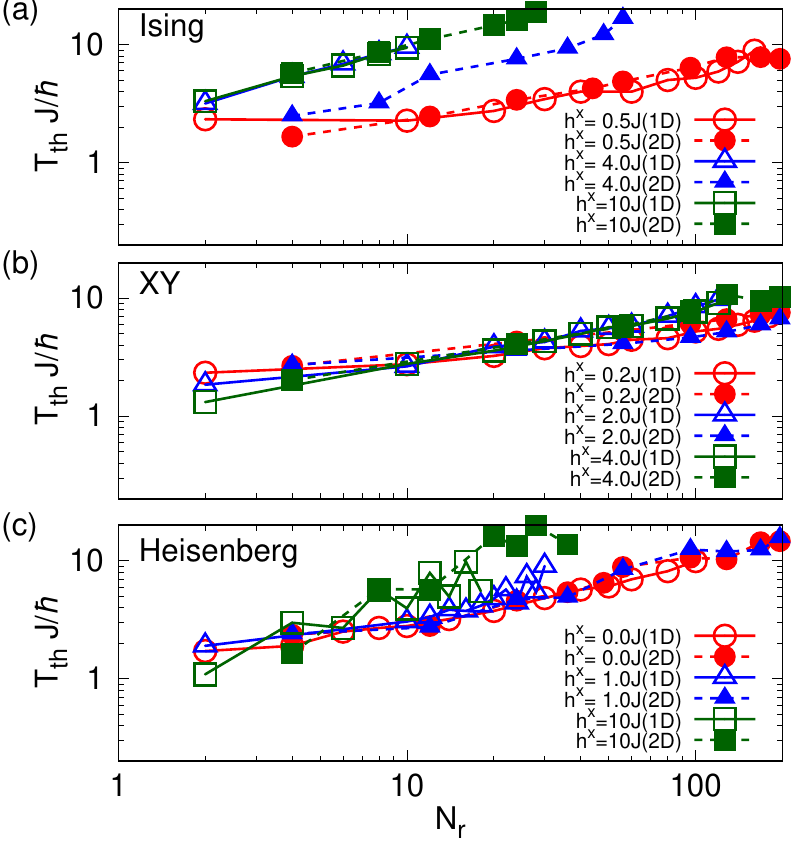}
\caption{Comparison with the threshold time for Ising model in 1D and 2D. (a) Ising model. (b) XY model. (c) Heisenberg model. Open and closed symbols represent 1D and 2D results, respectively.}
\label{fig:T_th_for_1D_2D}
\end{figure}%

Here, we compare the results of the threshold time in 1D and 2D. The results are shown in Fig.~\ref{fig:T_th_for_1D_2D} (a) for the Ising case, Fig.~\ref{fig:T_th_for_1D_2D} (b) for the XY case, and Fig.~\ref{fig:T_th_for_1D_2D} (c) for the Heisenberg case. From these results, we can see that the threshold time increases as power law of $N_r$ in all the models and spatial dimensions. We can also find that $T_{\rm th}$ results are almost overlapped in 1D and 2D at the same $h^x$ except the Ising model for $h^x=4.0J$, which will be discussed later. These results suggest that the threshold time depends on the number of interacting spins. We conclude that when the number of interacting spins is increased, the validity timescale of the DTWA becomes longer.

Here, we remark on the result of the Ising model for $h^x=4.0J$ (see blue triangle symbols in Fig.~\ref{fig:T_th_for_1D_2D}.). Unlike the other results, the 1D and 2D results are clearly deviated. This deviation may be attributed to the distance of the parameter from the quantum critical point. The critical field strength of the transverse field Ising model for $r=1$ is $h^x_{\rm c}=0.5J$ for 1D \cite{Sachdev_text_book} and $h_{\rm c}^x\simeq 3.044J$ for 2D \cite{Jongh1998,Rieger1998} in our notation. This means that the distance of the magnetic field from the quantum critical point is different in 1D and 2D. In 2D, $h_x=4.0J$ is closer to the critical value than in the 1D case. Therefore, the 1D and 2D results deviate clearly.


\section{Summary}\label{sec:summary}

In summary, we investigated the timescale on which the DTWA can quantitatively describe quantum dynamics of spin-$\frac{1}{2}$ models with a step function type spin-spin interaction in the short- and intermediate timescales. In order to corroborate this, we developed a formulation of the R\'enyi entropy within the DTWA framework. Using this formulation, we evaluate the R\'enyi entropy after a sudden quench in the Ising, XY, and Heisenberg models under a uniform magnetic field in 1D and 2D. The R\'enyi entropy is calculated by the DTWA and its extension including the second-order correction that is derived from the BBGKY hierarchy equation. Comparing these results, we determined the threshold time, on which the relative error of the exponential of the R\'enyi entropy in the DTWA and second-order BBGKY results exceeds $10\%$. We found that the threshold time increases as a power-law function of the interaction range (or the number of interacting spins per site). This result suggests that the accuracy of the DTWA becomes better as the classical limit (in this case, all-to-all coupling) is approached. This behavior is consistent with the properties of the equilibrium cases.

In this paper, we focused on the sudden quench dynamics of the quantum spin systems. The sweep dynamics, in which a parameter of the system varies slowly, is also an important problem. The adiabatic sweep of the parameter across the phase transition point leads the Kibble-Zurek mechanism. The DTWA can be applicable to this phenomenon. The adiabatic sweep is necessary for long-time evolution. Therefore, it is important to confirm the validity of the DTWA in the sweep dynamics case. 

\begin{acknowledgments}

The tDMRG calculations in this work are performed with ITensor library \cite{Fishman2007}. This work was supported by MEXT Quantum Leap Flagship Program (MEXT Q-LEAP) Grant Number JPMXS0118069021 (M.K. and I.D.), JST CREST No. JPMJCR1673 (I.D.), and JSPS KAKENHI Grant Number JP20K14389 (M.K.), JP20K14377 (S.G.), JP18K03492 (I.D.), and JP18H05228 (I.D.).

\end{acknowledgments}


\appendix
\section{Details of the derivation of the DTWA}\label{sec:details_of_DTWA}
In this appendix, we discuss the derivation of the DTWA for more details. See also Ref.~\cite{Pucci2016}.

A starting point for deriving the BBGKY hierarchy equation is the von Neumann equation for the density-matrix operator:
\begin{align}
i\hbar\frac{d}{d t}\hat{\rho}(t)=[\hat{H},\hat{\rho}(t)],\label{eq:von_Neumann_equation_for_density_matrix_operator}
\end{align}
where $\hat{\rho}(t)\equiv e^{-i\hat{H}t/\hbar}\hat{\rho}(0)e^{+i\hat{H}t/\hbar}$. By using the phase-point operator, the density matrix operator $\hat{\rho}(t)$ can be written as \cite{Wootters1987}
\begin{align}
\hat{\rho}(t)=\sum_{\bm{\alpha}}W_{\bm{\alpha}}(0)\hat{A}_{\bm{\alpha}}(t).\label{eq:expansion_density_matrix_by_phase_point_operator}
\end{align}
Substituting Eq.~(\ref{eq:expansion_density_matrix_by_phase_point_operator}) into Eq.~(\ref{eq:von_Neumann_equation_for_density_matrix_operator}), we obtain the equation for $\hat{A}_{\bm{\alpha}}(t)$:
\begin{align}
i\hbar\frac{d}{d t}\hat{A}_{\bm{\alpha}}(t)=[\hat{H},\hat{A}_{\bm{\alpha}}(t)].\label{eq:time_evolution_for_phase_point_operator}
\end{align}

Here, we introduce the partial trace of the phase-point operator
\begin{align}
\hat{A}_i(t, \bm{\alpha})&\equiv {\rm Tr}_i'\hat{A}_{\bm{\alpha}}(t),\label{eq:partial_trace_of_phase-point_operator_i}\\
\hat{A}_{i j}(t, \bm{\alpha})&\equiv {\rm Tr}_{i j}'\hat{A}_{\bm{\alpha}}(t),\label{eq:partial_trace_of_phase-point_operator_ij}
\end{align}
where ${\rm Tr}_i'$ and ${\rm Tr}_{i j}'$ are the trace over the Hilbert space except the site $i$ and sites $i$ and $j$ ($i\not=j$), respectively. To derive the BBGKY hierarchy equation, we use the following cluster expansion:
\begin{align}
\hat{A}_{ij}(t,\bm{\alpha})&\equiv\hat{A}_i(t, \bm{\alpha})\hat{A}_j(t,\bm{\alpha})+\hat{B}_{i j}(t,\bm{\alpha}),\label{eq:cluster_expansion_of_A_ij}\\
\hat{A}_{i j k}(t, \bm{\alpha})&\equiv \hat{A}_i(t, \bm{\alpha})\hat{A}_j(t,\bm{\alpha})\hat{A}_k(t,\bm{\alpha})\notag \\
&\quad +\hat{A}_i(t,\bm{\alpha})\hat{B}_{j k}(t,\bm{\alpha})+\hat{A}_j(t,\bm{\alpha})\hat{B}_{i k}(t,\bm{\alpha})\notag \\
&\quad +\hat{A}_k(t,\bm{\alpha})\hat{B}_{i j}(t,\bm{\alpha})+\hat{B}_{i j k}(t, \bm{\alpha}).\label{eq:cluster_expansion_Aijk}
\end{align}
$\hat{A}_i(t,\bm{\alpha})$ and $\hat{B}_{i j}(t, \bm{\alpha})$ can be expanded as
\begin{align}
\hat{A}_i(t, \bm{\alpha})&=\frac{1}{2}+\bm{r}_i(t,\bm{\alpha})\cdot\hat{\bm{S}}_i,\label{eq:expansion_partial_trace_Ai}\\
\hat{B}_{i j}(t, \bm{\alpha})&=4c_{i j}^{\mu\nu}(t,\bm{\alpha})\hat{S}_i^{\mu}\hat{S}_j^{\nu},\label{eq:expsnsion_partial_trace_B_ij}
\end{align}
where $\bm{r}_i(t,\bm{\alpha})$ and $c_{i j}^{\mu\nu}(t,\bm{\alpha})$ are expansion coefficients and determined by solving the classical equation of motion, which will be discussed below, and we use the Einstein's notation for Greek indices in Eq.~(\ref{eq:expsnsion_partial_trace_B_ij}). We note that the relation $c_{ij}^{\mu\nu}(t)=c_{ji}^{\nu\mu}(t)$ holds.

Setting $\hat{B}_{i j}(t, \bm{\alpha})=0$ in Eq.~(\ref{eq:cluster_expansion_of_A_ij}), we can obtain the first-order BBGKY hierarchy equation, which corresponds to the conventional DTWA. The Wigner-Weyl symbol of the spin operator $\hat{S}_i^{\mu}$ becomes 
\begin{align}
(\hat{S}_i^{\mu}(t))_{\rm W}={\rm Tr}[\hat{S}_i^{\mu}\hat{A}_{\bm{\alpha}}(t)]\simeq [\bm{r}_i(t,\bm{\alpha})]_{\mu}/2\equiv S_i^{\mu}(t).\label{eq:Wigner_Weyl_symbol_of_S}
\end{align}
The equation of motion for $S_i^{\mu}(t)$ can be obtained by the time derivative of Eq.~(\ref{eq:partial_trace_of_phase-point_operator_i}):
\begin{align}
\hbar\frac{d}{d t}S_i^{\mu}(t)&=\epsilon_{\mu\beta\gamma}\left[h^{\beta}S_i^{\gamma}(t)+\sum_{k\not=i}J_{i k}^{\beta}S_k^{\beta}(t)S_i^{\gamma}(t)\right],\label{eq:classical_equation_of_motion_DTWA_1st_app}
\end{align}
where we used Eqs.~(\ref{eq:time_evolution_for_phase_point_operator}) and (\ref{eq:cluster_expansion_of_A_ij}) and set $\hat{B}_{i j}(t,\bm{\alpha})=0$. 
We note that $c_{ij}^{\mu\nu}(t)=0$ in the 1st order BBGKY hierarchy equation because $\hat{B}_{i j}(t,\bm{\alpha})=0$. 

To derive the second-order BBGKY hierarchy equation, we remain $\hat{B}_{ij}(t, \bm{\alpha})$ and set $\hat{B}_{i j k}(t, \bm{\alpha})=0$. From the time derivative of Eqs.~(\ref{eq:partial_trace_of_phase-point_operator_i}) and (\ref{eq:cluster_expansion_of_A_ij}), we obtain
\begin{widetext}
\begin{align}
\hbar\frac{d}{d t}S_i^{\mu}(t)&=\epsilon_{\mu\beta\gamma}\left[h^{\beta}S_i^{\gamma}(t)+G_i^{\beta}(t)S_i^{\gamma}(t)+G_i^{\gamma\beta}(t)\right],\label{eq:BBGKY_first_order_equation_re}\\
\hbar\frac{d}{d t}c_{i j}^{\mu\nu}(t)&=\frac{1}{4}\epsilon_{\mu\nu\beta}\left[J_{i j}^{\nu}S_i^{\beta}(t)-J_{i j}^{\mu}S_j^{\beta}(t)\right]+\epsilon_{\beta\gamma\mu}h^{\beta}c_{ij}^{\gamma\nu}(t)+\epsilon_{\beta\gamma\nu}h^{\beta}c_{i j}^{\mu\gamma}(t)\notag \\
&\quad +\epsilon_{\beta\gamma\mu}G_{i\not{j}}^{\beta}(t)c_{i j}^{\gamma\nu}(t)+\epsilon_{\beta\gamma\nu}G_{j\not{i}}^{\beta}(t)c_{ij}^{\mu\gamma}(t)+\epsilon_{\beta\gamma\mu}S_i^{\gamma}(t)G_{i j}^{\nu\beta}(t)+\epsilon_{\beta\gamma\nu}S_j^{\gamma}(t)G_{j i}^{\mu\beta}(t)\notag \\
&\quad -\epsilon_{\beta\gamma\nu}J_{i j}^{\beta}S_i^{\mu}(t)\left[c_{i j}^{\beta\gamma}(t)+S_i^{\beta}(t)S_j^{\gamma}(t)\right]-\epsilon_{\beta\gamma\mu}J_{i j}^{\beta}S_j^{\nu}(t)\left[c_{ij}^{\gamma\beta}(t)+S_i^{\gamma}(t)S_j^{\beta}(t)\right],\label{eq:BBGKY_second_order_re}
\end{align}
\end{widetext}
where we used Eqs.~(\ref{eq:time_evolution_for_phase_point_operator}) and (\ref{eq:cluster_expansion_Aijk}),  set $\hat{B}_{i j k}(t,\bm{\alpha})=0$, and defined
\begin{align}
G^{\mu}_i(t)&\equiv \sum_{j\not=i}J_{i j}^{\mu}S^{\mu}_j(t),\label{eq:definition_of_Gz_re}\\
G_i^{\mu\nu}(t)&\equiv \sum_{j\not=i}J_{i j}^{\nu} c_{j i}^{\nu\mu}(t),\label{eq:definition_G_munu_re}\\
G^{\mu}_{i \not{j}}(t)&\equiv \sum_{k\not=i,j}J_{i k}^{\mu}S^{\mu}_k(t),\label{eq:definition_of_G_ij_z_re}\\
G_{i j}^{\mu\nu}(t)&\equiv \sum_{k\not=i,j}J_{i k}^{\nu}c_{j k}^{\mu \nu}(t).\label{eq:definition_of_G_ij_munu_re}
\end{align}

\section{Details of the sampling of the initial state}\label{sec:details_of_DTWA_sampling}

In this appendix, we discuss the sampling scheme of the initial states in the DTWA. Because initial conditions (\ref{eq:initial_state_for_Ising_model}), (\ref{eq:initial_state_for_XY_model}), and (\ref{eq:initial_state_for_Heisenberg_model}) are direct products states, we can write the density-matrix operator $\hat{\rho}(0)$ as a product of the density-matrix operator for each site $\hat{\rho}_i(0)$:
\begin{align}
\hat{\rho}(0)=\prod_{i=1}^M\hat{\rho}_i(0).\label{eq:product_of_density_matrix_operator_initial_states}
\end{align}
From Eq.~(\ref{eq:product_of_density_matrix_operator_initial_states}), we can also write the discrete Wigner function as a product of the discrete Wigner function for each site: $W_{\bm{\alpha}}(0)\equiv \prod_{i=1}^Mw_{\alpha_i}(0)$ and $w_{\alpha_i}(0)\equiv {\rm Tr}[\hat{\rho}_i\hat{A}_{\bm{\alpha}}(0)]$. The discrete Wigner functions for $\cket{\uparrow}_i$, $\cket{\downarrow_i}$, and $\cket{\leftarrow_i}$ are given by
\begin{align}
w_{\alpha_i}(0)=
\begin{cases}
1/2,\text{ for }\alpha_i=(0,0),(0,1),\\
0,\text{ for }\alpha_i=(1,0), (1,1),
\end{cases}
\text{for }\cket{\uparrow_i},\label{eq:discrete_Wigner_function_up_state}\\
w_{\alpha_i}(0)=
\begin{cases}
1/2,\text{ for }\alpha_i=(1,0),(1,1),\\
0,\text{ for }\alpha_i=(0,0), (0,1),
\end{cases}
\text{for }\cket{\downarrow_i},\label{eq:discrete_Wigner_function_down_state}\\
w_{\alpha_i}(0)=
\begin{cases}
1/2,\text{ for }\alpha_i=(0,1),(1,1),\\
0,\text{ for }\alpha_i=(0,0), (1,0),
\end{cases}
\text{for }\cket{\leftarrow_i}.\label{eq:discrete_Wigner_function_left_state}
\end{align}
The above discrete Winger functions are semi-positive definite and normalized: $\sum_{\alpha_i}w_{\alpha_i}(0)=1$. Therefore, we can regard the above discrete Wigner functions as a probability distribution function and use the Monte Carlo sampling for the initial states.

The initial value of the classical variable $S_i^{\mu}(0)$ is determined by Eq.~(\ref{eq:Wigner_Weyl_symbol_of_S}): $S_i^{\mu}(0)=[\bm{r}_i(0,\bm{\alpha})]_{\mu}/2=[\bm{r}_{\alpha_i}]_{\mu}/2$, where $\alpha_i$ is sampled from the discrete Wigner function and $[\bm{r}_{\alpha_i}]_{\mu}$ denotes $\mu$ component of the vector $\bm{r}_{\alpha_i}$. For example, we choose $\bm{S}_i(0)=(1,1,1)/2$ or $(-1,-1,1)/2$ for $\cket{\uparrow_i}$ state with the probability $\frac{1}{2}$ [see Eq.~(\ref{eq:discrete_Wigner_function_up_state})]. We note that $c_{i j}^{\mu\nu}(0)=0$ because the initial states are products states.

As pointed out in Ref.~\cite{Pucci2016}, there is ambiguity of the sampling scheme of the initial states because we have degrees of freedom of the definition of the phase-point operator (or definition of $\bm{r}_{\alpha}$). For example, if we use two different phase-point operators, we can decompose the density-matrix operator as $\hat{\rho}(0)=\sum_{\bm{\alpha}}[W_{\bm{\alpha}}(0)\hat{A}_{\bm{\alpha}}/2+W'_{\bm{\alpha}}(0)\hat{A}'_{\bm{\alpha}}/2]$, where $W_{\bm{\alpha}}'(0)$ and $\hat{A}_{\bm{\alpha}}'$ are the discrete Wigner function and phase-point operator for different definitions. This means that we have infinite number of possible choices of the sampling. According to Ref.~\cite{Pucci2016}, the suitable choice of the phase-point operator depends on the model and initial condition and they proposed some better choices rather than using Eqs.~(\ref{eq:discrete_Wigner_function_up_state}), (\ref{eq:discrete_Wigner_function_down_state}), and (\ref{eq:discrete_Wigner_function_left_state}). To implement the modified sampling, we introduce the following quantities:
\begin{align}
\bm{r}'_{\alpha}&\equiv ((-1)^{\alpha_2}, (-1)^{1+\alpha_1+\alpha_2}, (-1)^{\alpha_1}),\label{eq:definition_of_r_prime}\\
\tilde{\bm{r}}_{\alpha}&\equiv (\bm{r}_{\alpha}+\bm{r}_{\alpha}')/2,\label{eq:definition_of_tilde_r}\\
\bm{r}''_{\alpha}&\equiv ((-1)^{1+\alpha_2}, (-1)^{\alpha_1+\alpha_2}, (-1)^{\alpha_1}),\label{eq:definition_r_prime_prime}\\
\tilde{\bm{r}}_{\alpha}'&\equiv (\bm{r}_{\alpha}+\bm{r}_{\alpha}'')/2.\label{eq:definition_r_prime_tilde}
\end{align}
For the Ising and XY models, we sample $\bm{r}_{\alpha_i}$ from the following set with equal probability \cite{Pucci2016}:
\begin{align}
&S_{\rm Ising}\notag \\
&=\{\bm{r}_{(0,1)},\bm{r}_{(1,1)},\bm{r}'_{(0,1)},\bm{r}'_{(1,1)},\tilde{\bm{r}}_{(0,1)},\tilde{\bm{r}}_{(1,1)},\tilde{\bm{r}}'_{(0,1)},\tilde{\bm{r}}'_{(1,1)}\}.\label{eq:set_of_Ising_sampling}\\
&S_{\rm XY}=\{\tilde{\bm{r}}_{(1,0)},\tilde{\bm{r}}_{(1,1)},\tilde{\bm{r}}'_{(1,0)},\tilde{\bm{r}}'_{(1,1)}\}.\label{eq:set_of_XY_sampling}
\end{align}
For the Heisenberg model, we have checked that the following set gives better results:
\begin{align}
S_{\rm Heisenberg}=
\begin{cases}
\{\bm{r}_{(0,0)}, \bm{r}_{(0,1)}, \bm{r}'_{(0,0)}, \bm{r}'_{(0,1)}\},\quad \text{for }\cket{\uparrow_i},\\
\{\bm{r}_{(1,0)}, \bm{r}_{(1,1)}, \bm{r}'_{(1,0)}, \bm{r}'_{(1,1)}\},\quad \text{for }\cket{\downarrow_i}.
\end{cases}
\label{eq:set_of_Heisenberg_model}
\end{align}

\section{Derivation of the expression of R\'{e}nyi entropy}\label{sec:derivation_of_Renyi_entropy}

In this appendix, we derive the expression of the R\'{e}nyi entropy in the DTWA. Here, we consider the subsystem $A$ and its complement $B$. The total system is given by sum of the $A$ and $B$. The reduced density matrix for subsystem $A$ is defined by
\begin{align}
\hat{\rho}_A(t)\equiv {\rm Tr}_B[\hat{\rho}(t)],\label{eq:definition_of_reduced_density_matrix_app}
\end{align}
where ${\rm Tr}_B$ denotes the trace over the subsystem $B$. The second-order R\'{e}nyi entropy is defined by
\begin{align}
S^{(2)}_A(t)\equiv -\log\left({\rm Tr}\left\{[\hat{\rho}_A(t)]^2\right\}\right).\label{eq:definition_of_second_order_Renyi_entropy_app}
\end{align}
Using the discrete Wigner function [see Eq.~(\ref{eq:expansion_density_matrix_by_phase_point_operator})], we can write Eq.~(\ref{eq:definition_of_reduced_density_matrix_app}) as
\begin{align}
\hat{\rho}_A(t)={\rm Tr}_B\sum_{\bm{\alpha}}W_{\bm{\alpha}}(0)\hat{A}_{\bm{\alpha}}(t).\label{eq:reduced_density_matrix_by_Discrete_Wigner_function_app}
\end{align}

In the first-order BBGKY hierarchy equation, we approximate the phase-point operator as $\hat{A}_{\bm{\alpha}}(t)\simeq \prod_{i=1}^M\hat{A}_{i}(t,\bm{\alpha})$. Substituting this expression into Eq.~(\ref{eq:reduced_density_matrix_by_Discrete_Wigner_function}), we obtain
\begin{align}
\hat{\rho}_A(t)\simeq \sum_{\bm{\alpha}}W_{\bm{\alpha}}(0)\prod_{i\in A}\hat{A}_{i}(t,\bm{\alpha}),\label{eq:expression_reduced_density_matrix_in_1st_order}
\end{align}
where we used ${\rm Tr}_i\hat{A}_{\alpha_i}(t)=1$. To obtain the R\'{e}nyi entropy, we calculate 
\begin{align}
{\rm Tr}[\hat{\rho}_A(t)]^2&= {\rm Tr}\sum_{\bm{\alpha},\bm{\alpha}'}W_{\bm{\alpha}}(0)W_{\bm{\alpha}'}(0)\prod_{i\in A}\hat{A}_{i}(t,\bm{\alpha})\hat{A}_{i}(t,\bm{\alpha}')\notag \\
&=\sum_{\bm{\alpha},\bm{\alpha}'}W_{\bm{\alpha}}(0)W_{\bm{\alpha}'}(0)\prod_{i\in A}\left[\frac{1}{2}+2\bm{S}_i(t)\cdot\bm{S}'_i(t)\right],\label{eq:expression_for_reduced_density_matrix_squared}
\end{align}
where the initial conditions of $\bm{S}_i(t)$ and $\bm{S}_i'(t)$ are sampled from $W_{\bm{\alpha}}(0)$ and $W_{\bm{\alpha}'}(0)$, respectively.  This expression implies that we can obtain the second-order R\'{e}nyi entropy by using the replica method. The procedure is as follows: First, we prepare two independent copies of the initial states and calculate the time evolution for two copies independently. Then, we calculate the ensemble average of $\prod_{i\in A}[1/2+2\bm{S}_i(t)\cdot\bm{S}_i'(t)]$. The second order R\'{e}nyi entropy in the first-order BBGKY is given by
\begin{align}
S_{A}^{(2)}(t)=-\log\left\langle\left\langle\prod_{i\in A}\left[\frac{1}{2}+2\bm{S}_i(t)\cdot\bm{S}'_i(t)\right]\right\rangle\right\rangle.\label{eq:1st_order_BBGKY_renyi_expression_app}
\end{align}

Next, we derive the expression of the R\'{e}nyi entropy in the second-order BBGKY hierarchy. On the contrary to the first-order BBGKY hierarchy, we restrict the subsystem size to two sites. This is due to a practical reason. In the second-order BBGKY, we need to approximate the phase-point operator $\hat{A}_{\bm{\alpha}}(t)$ by using the cluster expansion. If we consider a large subsystem, we need expressions for a higher-order cluster expansion of $\hat{A}_{\bm{\alpha}}(t)$, which is difficult to write. Therefore, we only consider the two-site R\'{e}nyi entropy $S_{ij}^{(2)}(t)$.

The reduced density-matrix operator in the second-order BBGKY becomes
\begin{align}
\hat{\rho}_{i j}(t)&\equiv {\rm Tr}_{i j}'[\hat{\rho}(t)]\notag \\
&\simeq \sum_{\bm{\alpha}}W_{\bm{\alpha}}(0)\left[\hat{A}_i(t, \bm{\alpha})\hat{A}_j(t, \bm{\alpha})+\hat{B}_{ij}(t,\bm{\alpha})\right],\label{eq:expression_two-site_reduced_density_matrix_operator_2nd_order}
\end{align}
where we used Eq.~(\ref{eq:cluster_expansion_of_A_ij}). Using this expression, we obtain
\begin{align}
{\rm Tr}[\hat{\rho}_{i j}(t)]^2&=\sum_{\bm{\alpha},\bm{\alpha}'}W_{\bm{\alpha}}(0)W_{\bm{\alpha}'}(0)\notag \\
&\quad \times \left\{\prod_{l=i,j}\left[\frac{1}{2}+2\bm{S}_l(t)\cdot\bm{S}'_l(t)\right]\right.\notag \\
&\quad\quad +c_{i j}'^{\mu\nu}(t)S_i^{\mu}(t)S_j^{\nu}(t)+c_{i j}^{\mu\nu}(t)S_i'^{\mu}(t)S_j'^{\nu}(t)\notag \\
&\quad\quad  \quad +c_{i j}^{\mu\nu}(t)c_{i j}'^{\mu\nu}(t)\Bigg\},\label{eq:expression_reduced_density_matrix_squared_2nd_order}
\end{align}
where the initial conditions for $c_{ij}^{\mu\nu}(t)$ and $c_{i j}'^{\mu\nu}(t)$ are sampled from $W_{\bm{\alpha}}(0)$ and $W_{\bm{\alpha}'}(0)$, respectively, and we also use the Einstein's notation for Greek indices. From Eq.~(\ref{eq:expression_reduced_density_matrix_squared_2nd_order}), we can obtain the two-site R\'enyi entropy in the second-order BBGKY hierarchy.

\section{Details of tDMRG calculations}\label{sec:comparison_with_DTWA_and_tDMRG}

For the tDMRG calculations shown in Sec.~\ref{subsec:criterion}, we use the optimized Forest-Ruth-type fourth order decomposition~\cite{omelyan_optimized_2002} and set time step \(\delta t\) to \(0.05 \hbar / J\). The truncation error is set to be \(10^{-10}\), and bond dimensions of MPS are allowed to increase up to 4000. Simulations based on MPS are efficient for spatially 1D system or low-entangled states. Thus, we can compare DTWA and tDMRG without difficulty when \(r\) is close to one or comparable to the system size.

Because of the SU(2) symmetry, the time evolution of the two-site R\'enyi entropy does not depend on the magnetic field \(h^x\) in the Heisenberg model. In order to utilize the Abelian symmetry for numerical efficiency, we set \(h^x\) to 0 in the tDMRG simulations.

\section{Time evolution of matrix product states with long-range interactions}\label{sec:new_algorithm}
If the Hamiltonian consists of two-site operators, bond terms, the time evolution of MPS can be performed by operating Trotter gates to MPS~\cite{vidal_efficient_2003,vidal_efficient_2004,white_real-time_2004,daley_time-dependent_2004}.
Even though the Hamiltonian has long-range interactions, one can perform the time evolution of MPS with utilizing the swap gates~\cite{stoudenmire_minimally_2010}.
It should be noted that the swap gates to be operated are not unique, and that even the number of required swap operations can be different.
Less swap operations require less computational resources.
One may come up with a good choice of swap gates if the types of bond terms are limited likewise Bauernfeind \textit{et al. }\cite{bauernfeind_comparison_2019}.
If the Hamiltonian consists of many types of bond terms likewise the long-range models such as Eq.~\eqref{eq:system_Hamiltonian}, finding out a good choice is quite an exhausting task.
In this appendix, we present an algorithm which automatically produces an efficient (maybe not best) choice of swap operations.

The Hamiltonian consisting of two-site operators can be expressed as
\begin{align}
    \hat{H} = \sum_{i < j} \hat{H}_{i,j},
\end{align}
and one can compute the Trotter gates \(\exp(-\mathrm{i}\delta t \hat{H}_{i,j})\) from bond terms \(\hat{H}_{i,j}\).
At first step, we group the pair indices of bond terms \([i, j]\) so that bond terms in the same group commute each other.
We also try to group bond terms with the same distance \(j-i\) and order groups in ascending order of the distance.
The grouping can be accomplished by Algorithm~\ref{alg:group}~\footnote{For fermionic systems, Algorithm~\ref{alg:group} works if every bond term contains only even product of fermionic operators.}.

\begin{figure}
\begin{algorithm}[H]
    \raggedright
    \caption{Group bond terms}
    \label{alg:group}
    \textbf{Inputs} \\
    List of pairs in bond terms: \(Bonds = [[i, j], \ldots]\) \\
    Number of lattice sites: \(M\) \\
    \textbf{Output} \\
    List of grouped bond terms: \(GroupList\) \\

    \begin{algorithmic}[1]
        \State Sort \(Bonds\) in ascending order of the first index of pairs \(i\)
        \State Sort \(Bonds\) stably in descending order of difference between the second and first indices of pairs \(j-i\)
        \State \(GroupList \gets []\)(empty list)
        \While{length of \(Bonds > 0\)}
            \State \(Group \gets []\)
            \State \(Remain \gets [1,2, \ldots, M]\)
            \ForAll{\([i,j]\) in \(Bonds\)}
                \If{\(i \in Remain\) \(\land \) \(j \in Remain\)} 
                    \State Append \([i, j]\) to \(Group\)
                    \State Delete \(i\) and \(j\) from \(Remain\)
                    \State Delete \([i, j]\) from \(Bonds\)
                \EndIf
            \EndFor
            \State Append \(Group\) to \(GroupList\)
        \EndWhile
        \State Reverse the order of \(GroupList\)
    \end{algorithmic}
\end{algorithm}
\end{figure}

\begin{figure}
\begin{algorithm}[H]
    \raggedright
    \caption{Obtain swap gates by the gnome sort}
    \label{alg:swap}
    \textbf{Inputs} \\
    Present arrangement of site indices: \(Present\) \\
    Target arrangement of site indices: \(Target\) \\
    \textbf{Output} \\
    List of indices to be swapped: \(Swaps\)

    \begin{algorithmic}[1]
        \State \(M \gets \) length of \(Target\)
        \For{\(i \gets 1, M\)}
            \State \(Val[Target[i]] \gets i\)
        \EndFor
        \State \(Swaps \gets []\)
        \State \(Gnome \gets 2\)
        \While{\(Gnome \leq M\)}
            \If{\(Val[Present[Gnome-1]] \leq Val[Present[Gnome]]\)}
                \State \(Gnome \gets Gnome + 1\)
            \Else
                \State Swap \(Present[Gnome-1]\) and \(Present[Gnome]\)
                \State Append \((Gnome-1, Gnome)\) to \(Swaps\)
                \State \(Gnome \gets Gnome - 1\)
                \If{\(Gnome = 1\)}
                    \State \(Gnome \gets 2\)
                \EndIf
            \EndIf
        \EndWhile
    \end{algorithmic}
\end{algorithm}
\end{figure}

\begin{figure}[H]
    \centering
    \includegraphics[width=\linewidth]{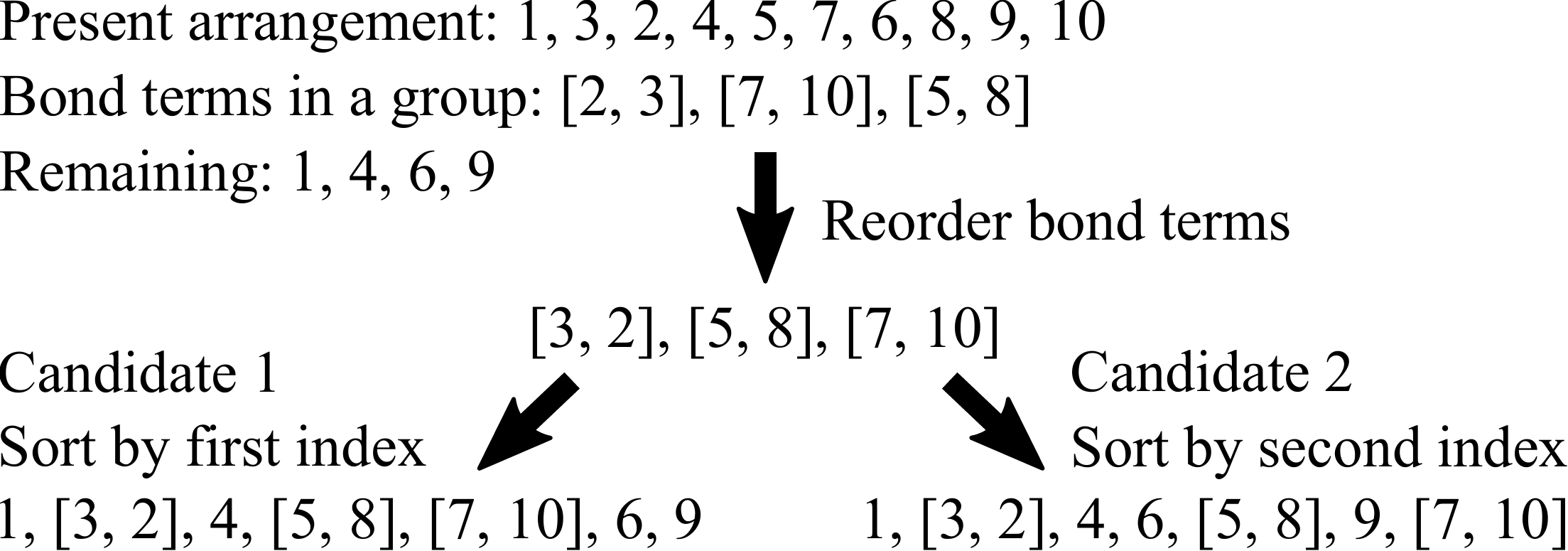}
    \caption{One example for obtaining two candidates of the arrangement of site indices. Both candidates can be obtained by four swap operations from the present arrangement.~\label{fig:example}}
\end{figure}

Next, we determine the arrangement of site indices where the Trotter gates in each group are operated. As shown in Algorithm~\ref{alg:swap}, swap operations required from one arrangement to another arrangement can be obtained by a sort algorithm implemented only by adjacent swap operations such as the bubble sort or the gnome sort. Therefore, it is sufficient to determine only the arrangement of site indices. In order to find an arrangement requiring less swap operations, we reorder bond terms in a group in ascending order in the sense of the present arrangement not to disturb the present arrangement so much. This ordering determines the arrangement of site indices in bond terms and we  have to insert remaining indices between bonds. Similarly to the case of the bond terms, ordering based on the present arrangement will not disturb the present arrangement so much. Here, we have a simple alternative: insert remaining indices based on the first index of bond pairs or the second one. From the alternative, we can obtain two candidates of an arrangement. The procedure for obtaining two candidates is summarized in Algorithm~\ref{alg:candidate}, and one example is given in Fig~\ref{fig:example}.

\begin{figure}
\begin{algorithm}[H]
    \raggedright
    \caption{Obtain candidate of arrangement}
    \label{alg:candidate}
    \textbf{Inputs} \\
    Present arrangement of site indices: \(Present\) \\
    Bond terms in a group: \(Bonds\) \\
    Index which determines candidate: \(k = 1\) or 2 \\
    \textbf{Output} \\
    Candidate arrangement: \(Candidate\) \\

    \begin{algorithmic}[1]
        \State \(M \gets \) length of \(Present\)
        \For{\(i \gets 1, M\)}
            \State \(Val[Present[i]] \gets i\)
        \EndFor
        \State \(Remain \gets [1, 2, \ldots, M]\)
        \ForAll{\([i,j]\) in \(Bonds\)}
            \State Delete \(i\) and \(j\) from \(Remain\)
            \If{\(Val[i] > Val[j]\)}
                \State Swap \(i\) and \(j\)
            \EndIf
        \EndFor
        \State Sort \(Bonds = [[i, j], \ldots]\) in ascending order by \(Val[i]\)
        \State Sort \(Remain = [i, \ldots]\) in ascending order by \(Val[i]\)
        \State \(Candidate \gets []\)
        \State \(Nr \gets \) length of \(Remain\)
        \State \(Nb \gets \) length of \(Bonds\)
        \State \(IndR \gets 1\)
        \State \(IndB \gets 1\)
        \While{\(IndR \leq Nr \lor IndB \leq Nb\)}
            \If{\(IndB > Nb \lor  (IndR \leq Nr \land Val[Remain[IndR]] < Val[Bonds[IndB][k]])\)}
                \State Append \(Remain[IndR]\) to \(Candidate\)
                \State \(IndR \gets IndR + 1\)
            \Else
                \State Append \(Bonds[IndB][1]\) to \(Candidate\)
                \State Append \(Bonds[IndB][2]\) to \(Candidate\)
                \State \(IndB \gets IndB + 1\)
            \EndIf
        \EndWhile
    \end{algorithmic}
\end{algorithm}
\end{figure}

From the two candidates, we select one candidate with considering possibilities in the next group.
With using Algorithm~\ref{alg:candidate}, one can obtain two candidates in the next group for each candidate.
Based on consequent four candidates, we choose one arrangement for the present group which is contained in the best candidates.
By iterating this process over groups, one can obtain a sequence of arrangement and swap operators required for performing time evolutions of long-range Hamiltonians.

The whole of above procedures is summarized in Algorithm~\ref{alg:find}.
Algorithm~\ref{alg:find} produces the efficient gates in Bauernfeind \textit{et al. }\cite{bauernfeind_comparison_2019} when \(bonds = [[1, 2], [1, 3], \ldots, [1, M]]\), and thus we consider that gates from the algorithm are efficient. 
From Trotter gates calculated from bond gates and the ordered list of swap operations given by Algorithm~\ref{alg:find}, one can obtain the list of gates corresponding to the first order decomposition of the time-evolution operator \(\prod_{i<j}\exp(-\mathrm{i}\delta t \hat{H}_{i, j})\).
The second order decomposition is obtained by successive operations of gates in the reversed list.
Furthermore, higher order decompositions can be obtained from compositions of the second order decompositions~\cite{omelyan_optimized_2002}.

\begin{figure*}
\begin{minipage}{\linewidth}
\begin{algorithm}[H]
    \raggedright
    \caption{Find efficient arrangements and swap operations}
    \label{alg:find}
    \textbf{Inputs} \\
    List of paris in bond terms: \(Bonds = [[i,j],\ldots]\) \\
    Number of lattice sites: \(M\) \\
    \textbf{Outputs} \\
    List of arrangements of site indices: \(Arrangements\) \\
    List of swap operations connecting arrangements: \(SwapsList\)

    \begin{algorithmic}[1]
        \State \(Arrange \gets [1, 2, \ldots, M]\)
        \State Get \(GroupList\) by Algorithm~\ref{alg:group} with inputs \((Bonds, M)\)
        \State \(Arrangements \gets []\)
        \State \(SwapsList \gets []\)
        \State \(L \gets\) length of \(GroupList\)
        \For{\(i=1, L-1\)}
            \State \(CandidateList \gets []\)
            \State \(NSwapsList \gets []\)
            \State \(SwapsCandidates \gets []\)
            \For{\(j=1, 2\)}
                \State Get \(Candidate\) by Algorithm~\ref{alg:candidate} with inputs \((Arrange, GroupList[i], j)\)
                \State Append \(Candidate\) to \(CandidateList\)
                \State Get \(Swaps\) by Algorithm~\ref{alg:swap} with inputs \((Arrange, Candidate)\)
                \State \(NSwaps \gets \) length of \(Swaps\)
                \State Append \(Swaps\) to \(SwapsCandidates\)
                \For{\(k=1,2\)}
                    \State Get \(NextCandidate\) by Algorithm~\ref{alg:candidate} with inputs \((Candidate, GroupList[i+1], k)\)
                    \State Get \(NextSwaps\) by Algorithm~\ref{alg:swap} with inputs \((Candidate, NextCandidate)\)
                    \State Append \(NSwaps +\)length of \(NextSwaps\)  to \(NSwapsList\) 
                \EndFor
            \EndFor
            \State \(MinSwaps \gets \) minimum of \(NSwapsList\) 
            \If{\(NSwapsList[1] = MinSwaps \lor NSwapsList[2] = MinSwaps\)}
                \State Append \(CandidateList[1]\) to \(Arrangements\)
                \State Append \(SwapsCandidates[1]\) to \(SwapsList\)
                \State \(Arrange \gets CandidateList[1]\)
            \Else
                \State Append \(CandidateList[2]\) to \(Arrangements\)
                \State Append \(SwapsCandidates[2]\) to \(SwapsList\)
                \State \(Arrange \gets CandidateList[2]\)
            \EndIf
        \EndFor
        \State Get \(Candidate1\) by Algorithm~\ref{alg:candidate} with inputs \((Arrange, GroupList[L], 1)\)
        \State Get \(Swaps1\) by Algorithm~\ref{alg:swap} with inputs \((Arrange, Candidate1)\)
        \State \(NSwaps1 \gets \) length of \(Swaps1\)
        \State Get \(Candidate2\) by Algorithm~\ref{alg:candidate} with inputs \((Arrange, GroupList[L], 2)\)
        \State Get \(Swaps2\) by Algorithm~\ref{alg:swap} with inputs \((Arrange, Candidate2)\)
        \State \(NSwaps2 \gets \) length of \(Swaps2\)
        \If{NSwaps1 < NSwaps2}
            \State Append \(Candidate1\) to \(Arrangements\)
            \State Append \(Swaps1\) to \(SwapsList\)
        \Else
            \State Append \(Candidate2\) to \(Arrangements\)
            \State Append \(Swaps2\) to \(SwapsList\)
        \EndIf
    \end{algorithmic}
\end{algorithm}
\end{minipage}
\end{figure*}

\clearpage

\begin{thebibliography}{999}
\bibitem{Bloch2012}
I. Bloch, J. Dalibard, and S. Nascimb\`ene, Quantum simulations with ultracold quantum gases, Nat. Phys. {\bf 8}, 267 (2012).
\bibitem{Trotzky2012}
S. Trotzky, Y.-A. Chen, A. Flesch, I, P. McCulloch, U. Schollw\"{o}ck, J. Eisert, and I. Bloch, Probing the relaxation towards equilibrium in an isolated strongly correlated one-dimensional Bose gas, Nat. Phys. {\bf 8}, 325 (2012).
\bibitem{Gring2012}
M. Gring, M. Kuhnert, T. Langen, T. Kitagawa, B. Rauer, M. Schreitl, I. Mazets, D. A. Smith, E. Demler, and J. Schmiedmayer, Relaxation and Prethermalization in an Isolated Quantum System, Science {\bf 337}, 1318 (2012).
\bibitem{Cheneau2012}
M. Cheneau, P. Barmettler, D. Poletti, M. Endres, P. Schau\ss, T. Fukuhara, C. Gross, I. Bloch, C. Kollath, and S. Kuhr, Light-cone-like spreading of correlations in a quantum many-body system, Nature {\bf 481}, 484 (2012).
\bibitem{Takasu2020a}
Y. Takasu, T. Yagami, H. Asada, Y. Fukushima, K. Nagao, S. Goto, I. Danshita, and Y. Takahashi, Energy redistribution and spatiotemporal evolution of correlations after a sudden quench of the Bose-Hubbard model, Sci. Adv. {\bf 6}, eaba9255 (2020).
\bibitem{Braun2015}
S. Braun, M. Friesdorf, S. S. Hodgman, M. Schriber, J. P. Ronzheimer, A. Riera, M. del Rey, I. Bloch, J. Eisert, and U. Schneider, Emergence of coherence and the dynamics of quantum phase transitions, PNAS {\bf 112}, 3641 (2015).
\bibitem{Choi2016}
J.-y. Choi, S. Hild, J. Zeiher, P. Schau\ss, A. Rubio-Abadal, T. Yefsah, V. Khemani, D. A. Huse, I. Bloch, and C. Gross, Exploring the many-body localization transition in two dimensions, Science {\bf 352}, 1547 (2016).
\bibitem{Abanin2019}
D. A. Abanin, E. Altman, I. Bloch, and M. Serbyn, Colloquium: Many-body localization, thermalization, and entanglement, Rev. Mod. Phys. {\bf 91}, 021001 (2019).
\bibitem{dePaz2013}
A. de Paz, A. Sharma, A. Chotia, E. Mar\'echal, J. H. Huckans, P. Pedri, L. Santos, O. Gorceix, L. Vernac, and B. Laburthe-Tolra, Nonequilibrium Quantum Magnetism in a Dipolar Lattice Gas, Phys. Rev. Lett. {\bf 111}, 185305 (2013).
\bibitem{Yan2013}
B. Yan, S. A. Moses, B. Gadway, J. P. Covey, K. R. A. Hazzard, A. M. Rey, D. S. Jin, and J. Ye, Observation of dipolar spin-exchange interactions with lattice-confined polar molecules, Nature {\bf 501}, 521 (2013).
\bibitem{Hazzard2014}
K. R. A. Hazzard, B. Gadway, M. Foss-Feig, B. Yan, S. A. Moses, J. P. Covey, N. Y. Yao, M. D. Lukin, J. Ye, D. S. Jin, and A. M. Rey, Many-Body Dynamics of Dipolar Molecules in an Optical Lattice, Phys. Rev. Lett. {\bf 113}, 195302 (2014).
\bibitem{dePaz2016}
A. de Paz, P. Pedri, A. Sharma, M. Efremov, B. Naylor, O. Gorceix, E. Mar\'echal, L. Vernac, and B. Laburthe-Tolra, Probing spin dynamics from the Mott insulating to the superfluid regime in a dipolar lattice gas, Phys. Rev. A {\bf 93}, 021603(R) (2016).
\bibitem{Lepoutre2019}
S. Lepoutre, J. Schachenmayer, L. Gabardos, B. Zhu, B. Naylor, E. Mar\'echal, O. Gorceix, A. M. Rey, L. Vernac, and B. Laburthe-Tolra, Out-of-equilibrium quantum magnetism and thermalization in a spin-3 many-body dipolar lattice system, Nat. Comm. {\bf 10}, 1714 (2019).
\bibitem{Fersterer2019}
P. Fersterer, A. Safavi-Naini, B. Zhu, L. Gabardos, S. Lepoutre, L. Vernac, B. Laburthe-Tolra, P. B. Blakie, and A. M. Rey, Dynamics of an itinerant spin-3 atomic dipolar gas in an optical lattice, Phys. Rev. A {\bf 100}, 033609 (2019).
\bibitem{Patscheider2020}
A. Patscheider, B. Zhu, L. Chomaz, D. Petter, S. Baier, A. M. Rey, F. Ferlaino, and M. J. Mark, Controlling dipolar exchange interactions in a dense three-dimensional array of large-spin fermions, Phys. Rev. Research {\bf 2}, 023050 (2020).
\bibitem{Schauss2015}
P. Schau\ss, J. Zeiher, T. Fukuhara, S. Hild, M. Cheneau, T. Macr\`i, T. Pohl, I. Bloch, and C. Gross, Crystallization in Ising quantum magnets, Science {\bf 347}, 1455 (2015).
\bibitem{Zeiher2016}
J. Zeiher, R. van Bijnen, P. Schau\ss, S. Hild, J.-y. Choi, T. Pohl, I. Bloch, and C. Gross, Many-body interferometry of a Rydberg-dressed spin lattice, Nat. Phys. {\bf 12}, 1095 (2016).
\bibitem{Labuhn2016}
H. Labuhn, D. Barredo, S. Ravets, S. de L\'es\'eleuc, T. Macr\`i, T. Lahaye, and A. Browaeys, Tunable two-dimensional arrays of single Rydberg atoms for realizing quantum Ising models, Nature {\bf 534}, 667 (2016).
\bibitem{Haenel2017}
R. Haenel, M. Schulz-Weiling, J. Sous, H. Sadeghi, M. Aghigh, L. Melo, J. S. Keller, and E. R. Grant, Arrested relaxation in an isolated molecular ultracold plasma, Phys. Rev. A {\bf 96}, 023613 (2017).
\bibitem{Zeiher2017}
J. Zeiher, J.-y. Choi, A. Rubio-Abadal, T. Pohl, R. van Bijnen, I. Bloch, and C. Gross, Coherent Many-Body Spin Dynamics in a Long-Range Interacting Ising Chain, Phys. Rev. X {\bf 7}, 041063 (2017).
\bibitem{Bernien2017}
H. Bernien, S. Schwartz, A. Keesling, H. Levine, A. Omran, H. Pichler, S. Choi, A. S. Zibrov, M. Endres, M. Greiner, V. Vuleti\'c, and M. D. Lukin, Probing many-body dynamics on a 51-atom quantum simulator, Nature {\bf 551}, 579 (2017).
\bibitem{Sous2018}
J. Sous and E. Grant, Possible Many-Body Localization in a Long-Lived Finite-Temperature Ultracold Quasineutral Molecular Plasma, Phys. Rev. Lett. {\bf 120}, 110601 (2018).
\bibitem{Leseleuc2018}
S. de L\'es\'eleuc, S. Weber, V. Lienhard, D. Barredo, H. P. B\"uchler, T. Lahaye, and A. Browaeys, Accurate Mapping of Multilevel Rydberg Atoms on Interacting Spin-1/2 Particles for the Quantum Simulation of Ising Models, Phys. Rev. Lett. {\bf 120}, 113602 (2018).
\bibitem{Guardado-Sanchez2018}
E. Guardado-Sanchez, P. T. Brown, D. Mitra, T. Devakul, D. A. Huse, P. Schau\ss, and W. S. Bakr, Probing the Quench Dynamics of Antiferromagnetic Correlations in a 2D Quantum Ising Spin System, Phys. Rev. X {\bf 8}, 021069 (2018).
\bibitem{Sous2019}
J. Sous and E. Grand, Many-body physics with ultracold plasmas: quenched randomness and localization, New J. Phys. {\bf 21}, 043033 (2019).
\bibitem{Leseleuc2019}
S. de L\'es\'eleuc, V. Lienhard, P. Scholl, D. Barredo, S. Weber, N. Lang, H. P. B\"uchler, T. Lahaye, and A. Browaeys, Observation of a symmetry-protected topological phase of interacting bosons with Rydberg atoms, Science {\bf 365}, 775 (2019).
\bibitem{Keesling2019}
A. Keesling, A. Omran, H. Levine, H. Bernien, H. Pichler, S. Choi, R. Samajdar, S. Schwartz, P. Silvi, S. Sachdev, P. Zoller, M. Endres, M. Greiner, V. Vulteti\'c, and M. D. Lukin, Quantum Kibble-Zurek mechanism and critical dynamics on a programmable Rydberg simulator, Nature {\bf 568}, 207 (2019).
\bibitem{Mizoguchi2019a}
M. Mizoguchi, Y. Zhang, M. Kunimi, A. Tanaka, S. Takeda, N. Takei, V. Bharti, K. Koyasu, T. Kishimoto, D. Jaksch, A. Glaetzle, M. Kiffner, G. Masella, G. Pupillo, M. Weidem\"uller, and K. Ohmori, Ultrafast Creation of Overlapping Rydberg Electrons in an Atomic BEC and Mott-Insulator Lattice, Phys. Rev. Lett. {\bf 124}, 253201 (2020)
\bibitem{Britton2012}
J. W. Britton, B. C. Sawyer, A. C. Keith, C.-C. J. Wang, J. K. Freericks, H. Uys, M. J. Biercuk, and J. J. Bollinger, Engineered two-dimensional Ising interactions in a trapped-ion quantum simulator with hundreds of spins, Nature {\bf 484}, 489 (2012).
\bibitem{Islam2013}
R. Islam, C. Senko, W. C. Campbell, S. Korenblit, J. Smith, A. Lee, E. E. Edwards, C.-C. J. Wang, J. K. Freericks, and C. Monroe, Emergence and Frustration of Magnetism with Variable-Range Interactions in a Quantum Simulator, Science {\bf 340}, 583 (2013).
\bibitem{Jurcevic2014}
P. Jurcevic, B. P. Lanyon, P. Hauke, C. Hempel, P. Zoller, R. Blatt, and C. F. Roos, Quasiparticle engineering and entanglement propagation in a quantum many-body system, Nature {\bf 511}, 202 (2014).
\bibitem{Richerme2014}
P. Richerme, Z.-X. Gong, A. Lee, C. Senko, J. Smith, M. Foss-Feig, S. Michalakis, A. V. Gorshkov, and C. Monroe, Non-local propagation of correlations in quantum systems with long-range interactions, Nature {\bf 511}, 198 (2014).
\bibitem{Bohnet2016}
J. G. Bohnet, B. C. Sawyer, J. W. Britton, M. L. Wall, A. M. Rey, M. Foss-Feig, and J. J. Bollinger, Quantum spin dynamics and entanglement generation with hundreds of trapped ions, Science {\bf 352}, 1297 (2016).
\bibitem{Garttner2017}
M. G\"arttner, J. G. Bohnet, A. Safavi-Naini, M. L. Wall, J. J. Bollinger, and A. M. Rey, Measuring out-of-time-order correlations and multiple quantum spectra in a trapped-ion quantum magnet, Nat. Phys. {\bf 13}, 781 (2017).
\bibitem{vidal_efficient_2003}
G. Vidal, Efficient Classical Simulation of Slightly Entangled Quantum Computations,  Phys. Rev. Lett. {\bf 91}, 147902 (2003).
\bibitem{white_real-time_2004}
S. R. White and A. E. Feiguin, Real-Time Evolution Using the Density Matrix Renormalization Group, Phys. Rev. Lett. {\bf 93}, 076401 (2004).
\bibitem{Blakie2008}
P. B. Blakie, A. S. Bradley, M. J. Davis, R. J. Ballagh, and C. W. Gardiner, Dynamics and statistical mechanics of ultra-cold Bose gases using c-field techniques,  Adv. Phys.{\bf 57}, 363 (2008).
\bibitem{Polkovnikov2010}
A. Polkovnikov, Phase space representation of quantum dynamics, Ann. Phys. {\bf 325}, 1790 (2010).
\bibitem{Tuchman2006}
A. K. Tuchman, C. Orzel, A. Polkovnikov, and M. A. Kasevich, Nonequilibrium coherence dynamics of a soft boson lattice, Phys. Rev. A {\bf 74}, 051601(R) (2006).
\bibitem{Davidson_thesis}
S. M. Davidson, Novel phase-space methods to simulate strongly-interacting many-body quantum dynamics, Doctor thesis (Boston university, 2017).
\bibitem{Nagao2019}
K. Nagao, M. Kunimi, Y. Takasu, Y.Takahashi, and I. Danshita, Semiclassical quench dynamics of Bose gases in optical lattices,  Phys. Rev. A {\bf 99}, 023622 (2019).
\bibitem{Ruostekoski2005}
J. Ruostekoski and L. Isella, Dissipative Quantum Dynamics of Bosonic Atoms in a Shallow 1D Optical Lattice, Phys. Rev. Lett. {\bf 95}, 110403 (2005).
\bibitem{Fujimoto2019}
K. Fujimoto, R. Hamazaki, and M. Ueda, Flemish Strings of Magnetic Solitons and a Nonthermal Fixed Point in a One-Dimensional Antiferromagnetic Spin-1 Bose Gas,  Phys. Rev. Lett. {\bf 122}, 173001 (2019).
\bibitem{Davidson2017}
S. M. Davidson, D. Sels, and A. Polkovnikov, Semiclassical approach to dynamics of interacting fermions, Ann. Phys. {\bf 384}, 128 (2017).
\bibitem{Polkovnikov2003}
A. Polkovnikov, Quantum corrections to the dynamics of interacting bosons: Beyond the truncated Wigner approximation, Phys. Rev. A {\bf 68}, 053604 (2003).
\bibitem{Davidson2015}
S. M. Davidson and A. Polkovnikov, SU(3) Semiclassical Representation of Quantum Dynamics of Interacting Spins, Phys. Rev. Lett. {\bf 114}, 045701 (2015).
\bibitem{Nagao2020a}
K. Nagao, Y. Takasu, Y. Takahashi, and I. Danshita, SU(3) truncated Wigner approximation for strongly interacting Bose gases, arXiv:2008.09900 [cond-mat.quant-gas] (2020).
\bibitem{Wurtz2018}
J. Wurtz, A. Polkovnikov, and D. Sels, Cluster truncated Wigner approximation in strongly interacting systems, Ann. Phys. {\bf 395}, 341 (2018).
\bibitem{Schmitt2019}
M. Schmitt, D. Sels, S. Kehrein, and A. Polkovnikov, Semiclassical echo dynamics in the Sachdev-Ye-Kitaev model, Phys. Rev. B {\bf 99}, 134301 (2019).
\bibitem{Scaffidi2019}
T. Scaffidi and E. Altman, Chaos in a classical limit of the Sachdev-Ye-Kitaev model, Phys. Rev. B {\bf 100}, 155128 (2019).
\bibitem{Schachenmayer2015}
J. Schachenmayer, A. Pikovski, and A. M. Rey, Many-Body Quantum Spin Dynamics with Monte Carlo Trajectories on a Discrete Phase Space, Phys. Rev. X {\bf 5}, 011022 (2015).
\bibitem{Schachenmayer2015_2}
J. Schachenmayer, A. Pikovski, and A. M. Rey, Dynamics of correlations in two-dimensional quantum spin models with long-range interactions: a phase-space Monte-Carlo study, New J. Phys. {\bf 17}, 065009 (2015).
\bibitem{Sundar2019}
B. Sundar, K. C. Wang, and K. R. A. Hazzard, Analysis of continuous and discrete Wigner approximations for spin dynamics, Phys. Rev. A {\bf 99}, 043627 (2019).
\bibitem{Orioli2018}
A. P. Orioli, A. Signoles, H. Wildhagen, G. G\"unter, J. Berges, S. Whitlock, and M. Weidem\"uller, Relaxation of an Isolated Dipolar-Interacting Rydberg Quantum Spin System, Phys. Rev. Lett. {\bf 120}, 063601 (2018).
\bibitem{Signoles2019a}
A. Signoles, T. Franz, R. A. Alves, M. G\"arttner, S. Whitlock, G. Z\"urn, and M. Weidem\"uller, Glassy dynamics in a disordered Heisenberg quantum spin system, arXiv:1909.11959 [quant-ph] (2019).
\bibitem{Babadi2015}
M. Babadi, E. Demler, and M. Knap, Far-from-Equilibrium Field Theory of Many-Body Quantum Spin Systems: Prethermalization and Relaxation of Spin Spiral States in Three Dimensions, Phys. Rev. X {\bf 5}, 041005 (2015).
\bibitem{Pucci2016}
L. Pucci, A. Roy, and M. Kastner, Simulation of quantum spin dynamics by phase space sampling of Bogoliubov-Born-Green-Kirkwood-Yvon trajectories, Phys. Rev. B {\bf 93}, 174302 (2016).
\bibitem{Acevedo2017}
O. L. Acevedo, A. Safavi-Naini, J. Schachenmayer, M. L. Wall, R. Nandkishore, and A. M. Rey, Exploring many-body localization and thermalization using semiclassical methods,  Phys. Rev. {\bf 96}, 033604 (2017).
\bibitem{Orioli2017}
A. Pi\~neiro Orioli, A. Safavi-Naini, M. L. Wall, and A. M. Rey, Nonequilibrium dynamics of spin-boson models from phase-space methods, Phys. Rev A {\bf 96}, 033607 (2017).
\bibitem{Czischek2018}
S. Czischek, M. G\"{a}rttner, M. Oberthaler, M. Kastner, and T. Gasenzer, Quenches near criticality of the quantum Ising chain-power and limitations of the discrete truncated Wigner approximation, Quantum Sci. Technol. {\bf 4}, 014006 (2018).
\bibitem{Covey2018}
J. P. Covey, L. De Marco, \'{O}. L. Acevedo, A. M. Rey, and J. Ye, An approach to spin-resolved molecular gas microscopy, New J. Phys. {\bf 20}, 043031 (2018).
\bibitem{Mori2019}
T. Mori, Prethermalization in the transverse-field Ising chain with long-range interactions, J. Phys. A:Math. Theor. {\bf 52}, 054001 (2019).
\bibitem{Qu2019}
C. Qu and A. M. Rey, Spin squeezing and many-body dipolar dynamics in optical lattice clocks, Phys. Rev. A {\bf 100}, 041602(R) (2019).
\bibitem{Morong2020a}
W. Morong, S. R. Muleady, I. Kimchi, W. Xu, R. M. Nandkishore, A. M. Rey, and B. DeMarco, Disorder-controlled relaxation in a 3D Hubbard
model quantum simulator, arXiv:2001.07341 (2020). 
\bibitem{Zhu2019}
B. Zhu, A. M. Rey, and J. Schachenmayer, A generalized phase space approach for solving quantum spin dynamics, New J. Phys. {\bf 21}, 082001 (2019).
\bibitem{Henkel2010}
N. Henkel, R. Nath, and T. Pohl, Three-Dimensional Roton Excitations and Supersolid Formation in Rydberg-Excited Bose-Einstein Condensates, Phys. Rev. Lett. {\bf 104}, 195302 (2010).
\bibitem{Pupillo2010}
G. Pupillo, A. Micheli, M. Boninsegni, I. Lesanovsky, and P. Zoller, Strongly Correlated Gases of Rydberg-Dressed Atoms: Quantum and Classical Dynamics, Phys. Rev. Lett. {\bf 104}, 223002 (2010).
\bibitem{Schachenmayer2013}
J. Schachenmayer, B. P. Lanyon, C. F. Roos, and A. J. Daley, Entanglement Growth in Quench Dynamics with Variable Range Interactions, Phys. Rev. X {\bf 3}, 031015 (2013).
\bibitem{Hauke2013}
P. Hauke and L. Tagliacozzo, Spread of Correlations in Long-Range Interacting Quantum Systems, Phys. Rev. Lett. {\bf 111}, 207202 (2013).
\bibitem{Lepori2016}
L. Lepori, A. Trombettoni, and D. Vodola, Singular dynamics and emergence of nonlocality in long-range quantum models, J. Stat. Mech. 033102 (2017).
\bibitem{Buyskikh2016}
A. S. Buyskikh, M. Fagotti, J. Schachenmayer, F. Essler, and A. J. Daley, Entanglement growth and correlation spreading with variable-range interactions in spin and fermionic tunneling models, Phys. Rev. A {\bf 93}, 053620 (2016).
\bibitem{Wootters1987}
W. K. Wootters, A Wigner-Function Formulation of Finite-State Quantum Mechanics, Ann. Phys. {\bf 176}, 1 (1987).
\bibitem{Eisert2010}
J. Eisert, M. Cramer, and M. B. Plenio, Colloquium: Area laws for the entanglement entropy, Rev. Mod. Phys {\bf 82}, 277 (2010).
\bibitem{Vidal2004}
J. Vidal, G. Palacios, and C. Aslangul, Entanglement dynamics in the Lipkin-Meshkov-Glick model, Phys. Rev. A {\bf 70}, 062304 (2004).
\bibitem{Latorre2005}
J. I. Latorre, R. Or\'us, E. Rico, and J. Vidal, Entanglement entropy in the Lipkin-Meshkov-Glick model, Phys. Rev. A {\bf 71}, 064101 (2005).
\bibitem{Sachdev_text_book}
S. Sachdev, {\it Quantum Phase Transitions} (Cambridge University Press, Cambridge, England, 2011)
\bibitem{Jongh1998}
M. S. L. du Croo de Jongh and J. M. J. van Leeuwen, Critical behavior of the two-dimensional Ising model in a transverse field: A density-matrix renormalization calculation, Phys. Rev. B {\bf 57}, 8494 (1998).
\bibitem{Rieger1998}
H. Rieger and N. Kawashima, Application of a continuous time cluster algorithm to the two-dimensional random quantum Ising ferromagnet, Eur. Phys. J. B {\bf 9}, 233 (1999).
\bibitem{Fishman2007}
M. Fishman, S. R. White, and E. M. Stoudenmire, The ITensor Software Library for Tensor Network Calculations, arXiv:2007.14822 (2020).
\bibitem{omelyan_optimized_2002}
I. P. Omelyan, I. M. Mryglod, and R. Folk, Optimized Forest-Ruth- and Suzuki-like algorithms for integration of motion in many-body systems, Comput. Phys. Commun. {\bf 146}, 188 (2002).
\bibitem{vidal_efficient_2004}
G. Vidal, Efficient Simulation of One-Dimensional Quantum Many-Body Systems, Phys. Rev. Lett. {\bf 93}, 040502 (2004).
\bibitem{daley_time-dependent_2004}
A. J. Daley, C. Kollath, U. Schollw\"ock, and G. Vidal, Time-dependent density-matrix renormalization-group using adaptive effective Hilbert spaces, J. Stat. Mech. (2004) P04005.
\bibitem{stoudenmire_minimally_2010}
E. M. Stoudenmire and S. R. White, Minimally entangled typical thermal state algorithms, New J. Phys. {\bf 12}, 055026 (2010).
\bibitem{bauernfeind_comparison_2019}
D. Bauernfeind, M. Aichhorn, and H. G. Evertz, Comparison of MPS based real time evolution algorithms for Anderson Impurity Models, arXiv:1906.09077 (2019).




\end{thebibliography}

\end{document}